\documentclass[a4paper,onecolumn,11pt,accepted=2026-04-27]{quantumarticle} 
\pdfoutput=1

\usepackage[utf8]{inputenc}
\usepackage[english]{babel}
\usepackage[T1]{fontenc}
\usepackage{amsmath}
\usepackage{hyperref}
\usepackage{amssymb}
\usepackage{graphicx}
\usepackage{braket}
\usepackage{algorithm}
\usepackage{algpseudocode}
\usepackage{float} 
\usepackage[dvipsnames]{xcolor}
\usepackage{amsthm}
\usepackage{qcircuit}
\usepackage{hyperref}
\usepackage{xfrac}
\usepackage{subcaption}
\usepackage{paralist}
\usepackage[textsize=footnotesize]{todonotes}

\usepackage[capitalise]{cleveref}
\usepackage{parskip}
\usepackage{comment}
\usepackage{multirow}
\usepackage{wrapfig}
\usepackage{mathtools}
\usepackage{stmaryrd}

\makeatletter
\newcommand{\xMapsto}[2][]{\ext@arrow 0599{\Mapstofill@}{#1}{#2}}
\def\Mapstofill@{\arrowfill@{\Mapstochar\Relbar}\Relbar\Rightarrow}
\makeatother

\newtheorem{theorem}{Theorem}
\newtheorem{lemma}{Lemma}

\newcommand{\arrow}[1]{\xrightarrow[]{#1}}

\newcommand{\rank}[1]{\chi(#1)}
\newcommand{\UnitaryRank}[1]{\chi(#1)}
\newcommand{\MaxGadgetSize}{\mu}

\newcommand{\SGate}{\text{S}}
\newcommand{\HGate}{\text{H}}
\newcommand{\CXGate}{\text{CX}}
\newcommand{\XGate}{\text{X}}
\newcommand{\ZGate}{\text{Z}}

\newcommand{\TGate}{\text{T}}
\newcommand{\RzGate}{\text{R}_Z}
\newcommand{\RxGate}{\text{R}_X}

\newcommand{\CCZGate}{\text{CCZ}}
\newcommand{\CCXGate}{\text{CCX}}

\newcommand{\INIGate}{\text{INI}}
\newcommand{\ANDGate}{\text{AND}}
\newcommand{\ADDGate}{\text{ADD}}
\newcommand{\ORGate}{\text{OR}}
\newcommand{\QFTGate}{\text{QFT}}
\newcommand{\MULGate}{\text{MUL}}
\newcommand{\INCGate}{\text{INC}}

\newcommand{\NP}{\mathbf{NP}}
\newcommand{\PneqNP}{\mathbf{P} \neq \NP}
\newcommand{\PeqNP}{\mathbf{P} = \NP}

\newcommand{\OOM}{\color{OrangeRed}{OOM}}
\newcommand{\TO}{\color{OrangeRed}{TO}}
\newcommand{\s}[1]{\color{ForestGreen}{#1}} 

\newcommand{\impl}[1]{\widehat{#1}}
\newcommand{\GATE}{G}

\newcommand{\booleanMapping}{g}

\newcommand{\Paragraph}[1]{\smallskip\noindent{\bf #1}}
\newcommand\numberthis{\addtocounter{equation}{1}\tag{\theequation}}

\begin{document}

\title{Efficient Simulation of High-Level Quantum Gates}

\author{Adam Husted Kjelstrøm}
\email{husted@cs.au.dk}
\affiliation{Department of Computer Science, Aarhus University}
\orcid{0000-0002-1275-9933}

\author{Andreas Pavlogiannis}
\email{pavlogiannis@cs.au.dk}
\affiliation{Department of Computer Science, Aarhus University}
\orcid{0000-0002-8943-0722}

\author{Jaco van de Pol}
\email{jaco@cs.au.dk}
\affiliation{Department of Computer Science, Aarhus University}
\orcid{0000-0003-4305-0625}

\maketitle

\begin{abstract}
Quantum circuit simulation is paramount to the verification and optimization of quantum algorithms, and considerable research efforts have been made towards efficient simulators. 
While circuits often contain high-level gates such as oracles and multi-controlled $\XGate$ ($\text{C}^k \XGate$) gates, existing simulation methods require compilation to a low-level gate-set before simulation.
This, however, increases circuit size and incurs a considerable (typically exponential) overhead, even when the number of high-level gates is small.
Here we present a gadget-based simulator which simulates high-level gates directly, thereby allowing to avoid or reduce the blowup of compilation.
Our simulator uses a stabilizer decomposition of the magic state of non-stabilizer gates, with improvements in the rank of the magic state directly improving performance. 
We then proceed to establish a small stabilizer rank for a range of high-level gates that are common in various quantum algorithms.
Using these bounds in our simulator, we improve both the theoretical complexity of simulating circuits containing such gates, and the practical running time compared to standard simulators found in IBM's Qiskit Aer library. 
We also derive exponential lower-bounds for the stabilizer rank of some gates under common complexity-theoretic hypotheses. In certain cases, our lower-bounds are asymptotically tight on the exponent.
\end{abstract}

\section{Introduction} \label{sec:intro}
Testing software is the most standard approach to detecting errors~\cite{Lampropoulos2017}.
In the quantum setting, this is often done by simulating quantum circuits on classical hardware~\cite{Wang2021}.
Strong simulation provides the probability of observing some specific measurement of the qubits after applying the target circuit.
Remarkably, strong simulation can also be used to prove two quantum circuits equivalent~\cite{Peham2023_depth_optimal_synthesis_of_clifford_circuits},
which has immediate applications in circuit optimization~\cite{Wille2023optimal_cliff_synth_via_SAT}.

The importance of strong simulation has led to the development of a variety of simulation techniques, targeting efficiency.
Given a quantum circuit of $n$ qubits and $m$ gates, the simplest way is to store each quantum state explicitly, requiring  time and space that is exponential in $n$, which is prohibitive.
Feynman~\cite{feynman1982simulating_phyiscs_with_computers} proposed a recursive approach with time exponential in $m$, at the benefit of polynomial space.
Aaronson and Gottesman~\cite{Aaronson04sim_of_stabilizer_circuits} showed that stabilizer circuits can be simulated in $O(n^2 m)$ time and $O(n^2)$ space.
Although stabilizer circuits are not capable of arbitrary quantum computations, they are powerful enough to generate quantum effects such as superposition and entanglement.

The stabilizer simulator has been extended to universal quantum circuits of stabilizer, $\TGate$, and $\CCXGate$ gates ~\cite{Bravyi16trading_classical_and_q_resources, Bravyi2019simulating_q_circuits_by_low_rank_stabilizer_decomp} by substituting each $\TGate$ and $\CCXGate$ gate by a small circuit that applies the same effect, using a, so called, ``magic state'' in the process~\cite{Zhou_2000_gadgetize_T_gates_magic}.
Specifically for $t$ $\TGate$ gates, the simulator runs in $O(2^{0.396t} n^2 m)$ time and $O(t+n^2)$ space~\cite{Gosset2021improved_upper_bounds_stabilizer_rank}.

While $\TGate$ or $\CCXGate$ gates suffice to make stabilizer circuits universal for quantum computing, quantum circuits are generally composed from a much richer gate-set \cite{de_Veras2022_CVO_QRAM_quantum_algorithm, Harrow2009_quantum_algorithm_for_linear_systems_of_equations, Guo2022_Quantum_algorithms_for_anomaly_detection_using_amplitude_estimation,Li2023_quantum_algorithm_for_k_fold_cross_validation_for_nearest_neighbor_classification}, 
including multi-controlled Toffoli gates ($C^k \XGate$), arbitrary multi-controlled single-qubit unitaries ($C^k U$), and oracle gates, which are a standard means of providing quantum access to a function. In particular, given a Boolean function $\varphi\colon \{0,1\}^k \arrow{} \{0,1\}$, the oracle gate $C_{\varphi} U$ applies the single qubit unitary $U$ to the target qubit $k+1$ when the state of the control qubits $1,\ldots,k$ satisfies $\varphi$.
Oracle gates are used in Grover's algorithm \cite{Grover1996FastQuantumMechanicalAlgorithm}, the Deutsch-Josza algorithm \cite{DeutschJozsa1992RapidSolutionProblemsQuantumComputation}, 
the Bernstein-Vazirani algorithm \cite{BernsteinVazirani1993QuantumComplexityTheory}, 
Shor's algorithm \cite{Shor1994AlgorithmsForQuantumComputation}, 
and Simon's algorithm \cite{Simon1997OnThePowerOfQuantumComputation},
among others.

Circuits containing such high-level gates can still be simulated somewhat efficiently (at least in theory), by compiling the high-level gates to stabilizer+$\TGate$ equivalents, with active research on designing efficient compilation schemes~\cite{Zindorf2024efficient_implementation_multi_controlled_quantum,Vale2023decomposition_of_multi_controlled_special_unitary_single_qubit_gates}.
In particular for $C^k U$, these schemes convert each such gate into $\Theta(k)$ $\TGate$ gates.
However, since the number of $\TGate$ gates contributes exponentially to the running time, simulating the compiled circuits has been a challenge.
\emph{Can we avoid the overhead of compilation schemes? Is it possible to simulate high-level circuits more efficiently?}

\noindent \textbf{Theoretical contributions.} 
We answer these questions by providing an extension to the existing stabilizer-based quantum simulation theory, generalizing from $\TGate$ and $\CCXGate$ gates. 
In particular, assume we want to simulate a target circuit $C$ containing some high-level gates.
Our method is to ``implement'' each such high-level gate $\GATE$ by some stabilizer circuit and some stabilizer decomposition of its magic input state. 
More precisely, let $C$ be a circuit on $n$ qubits, and assume that each high-level gate $\GATE$ in it can be implemented by a stabilizer circuit $\impl{\GATE}$ of $\leq \MaxGadgetSize$ gates, applied to a magic state $\ket{\psi}$ that is decomposed as a weighted sum of $k$ stabilizer states, each initialized by $\leq \MaxGadgetSize$ gates. Note that $k\geq \chi(\ket{\psi})$, the (minimal) stabilizer rank of $\ket{\psi}$ \cite{Bravyi16trading_classical_and_q_resources}.
We generally restrict $\impl{\GATE}$ to using $O(n)$ ancillae.
We call such an implementation a $(\MaxGadgetSize,k)$-decomposition of $\GATE$.
The following theorem describes the complexity of simulating $C$, given such an implementation.

\smallskip
\begin{theorem}
\label{theorem:main_algorithm}
    Let $C$ be a circuit over $n$ qubits and $m$ gates 
    composed of stabilizer gates as well as $t$ non-stabilizer gates $\GATE_1, \GATE_2, \dots, \GATE_t$.
    Let $x\in \{0,1\}^n$ be a measurement outcome.
    Assume that each $\GATE_j$ (for $1\leq j\leq t$) has a $(\MaxGadgetSize_j,k_j)$-decomposition using $O(n)$ ancillae.
    Set $\chi = \prod_j k_j$, and $\MaxGadgetSize = \max_{j} \MaxGadgetSize_j$.
    Then we can compute $|\braket{x|C|0^n}|^2$, the probability of observing $x$, in time $O(\chi n^2 (m+ t \MaxGadgetSize))$. 
    The algorithm uses $O(\log \chi+n^2)$ space, and can be parallelized for a run-time of $O(n^2 (m+t \MaxGadgetSize))$ on $\chi$ threads.  
\end{theorem}

We note that the restriction to using $O(n)$ ancillae suffices for the high-level gates we consider in this work, and it simplifies the formulation since the cost of $O(n)$ ancillae is absorbed in time bound mentioned in \cref{theorem:main_algorithm}.

We will illustrate applications to some high-level gates below.
Compared to the standard process of compiling high-level gates to Clifford+$\TGate$ and then using existing simulation techniques, our bounds are smaller as long as the individual values $k_j$ and $\MaxGadgetSize_j$ are small.
This motivates identifying an implementation for the high-level gates of interest that keeps these quantities small.
To this end, we consider various multi-controlled and oracle gates $\GATE$, and find $(\MaxGadgetSize,k)$-decompositions for which we establish small upper bounds $k$ on the stabilizer rank, and gate size $\MaxGadgetSize=O(n)$.

Towards a bound on the stabilizer rank of oracle gates $\rank{C_\varphi U}$, for some Boolean predicate $\varphi\colon \{0,1\}^k \arrow{} \{0,1\}$ and single-qubit unitary $U$, we define the ``effectual'' state $\ket{E(\varphi)}$ of $\varphi$ as the sum of all basis states $\ket{x}$ for which $\varphi(x)$ is true:
\[
\ket{E(\varphi)} \propto \sum_{x\in \{0,1\}^k\colon \varphi(x)} \ket{x} \ .
\]
Here the ``proportional to'' notation $\propto$ denotes equality up to a scaling factor.
We derive a bound on $\rank{C_\varphi U}$ as a function of $\rank{\ket{E(\varphi)}}$, generally establishing \cref{theorem:gate_rank_from_E_rank}.

\smallskip
\begin{theorem}
    \label{theorem:gate_rank_from_E_rank}
    Let $\varphi$ be a Boolean predicate.
    Then the following properties hold for the stabilizer rank, where $\RzGate(\theta)$ and $\RxGate(\theta)$ are the standard rotation gates, and $U$ is an arbitrary single-qubit unitary.
    \begin{align}
        \rank{C_\varphi \RzGate(\theta)} &\leq \rank{\ket{E(\varphi)}}+1                   \label{eq:rank_C_phi_Rz_bound_E} \\
        \rank{C_\varphi \RxGate(\theta)}    &= \rank{C_\varphi \RzGate(\theta)}            \label{eq:rank_C_phi_Z_eq_rank_C_phi_X} \\   
        \rank{C_\varphi U} &\leq
            \begin{cases} 
                8,               & \text{if } \rank{\ket{E(\varphi)}}=1 \\  
                8\rank{\ket{E(\varphi)}}+8,  & \text{otherwise.}
            \end{cases}                                                         \label{eq:rank_C_phi_U}
    \end{align}
\end{theorem}

A simple strategy for bounding $\rank{\ket{E(\varphi)}}$ is to note that for all $x\in\{0,1\}^k$, $\ket{x}$ is a stabilizer state, giving $\rank{\ket{E(\varphi)}}\leq |\{x:\varphi(x)\}|\leq 2^k$.
However, since $\varphi$ is an arbitrary formula, we do not have non-trivial upper bounds on $\chi(\ket{E(\varphi)})$. 
Nonetheless, as predicates are often inductively defined over logical connectives such as conjunctions, disjunctions, and negations, we derive more refined bounds by following this structure.

\smallskip
\begin{lemma}
    \label{lemma:rank_E_logical_connectives}
    Let $\varphi_1$ and $\varphi_2$ be Boolean formulas.
    The following inequalities hold.
    \begin{align}
        \rank{\ket{E(\neg \varphi_1)}}            &\leq \rank{\ket{E(\varphi_1)}} + 1                                     \label{eq:rank_E_negation} \\
        \rank{\ket{E(\varphi_1\wedge \varphi_2)}} &\leq \rank{\ket{E(\varphi_1)}}\rank{\ket{E(\varphi_2)}}     \label{eq:rank_E_conj_bound} \\
        \rank{\ket{E(\varphi_1\vee \varphi_2)}}   &\leq \rank{\ket{E(\varphi_1 \wedge\varphi_2)}} +\rank{\ket{E(\varphi_1)}}+\rank{\ket{E(\varphi_2)}}        \label{eq:rank_E_disj_bound} 
    \end{align}
\end{lemma}
Note that $|\{x:\varphi(x)\}|+|\{x:\neg\varphi(x)\}|=2^k$, which, together with \cref{eq:rank_E_negation}, implies $\rank{\ket{E(\varphi)}}\leq 2^{k-1}$ in general.
Finally, we establish some direct bounds when $\varphi$ has a specific form, in particular performing comparison between integers.

\smallskip
\begin{theorem}
    \label{theorem:rank_E_for_comparisons}
    Let $x$ and $y$ be formal variables ranging over $k$-bit integers.
    The following (in)equalities hold.
    \begin{align}
        \rank{\ket{E(x = y)}}     &= 1                                        \label{eq:rank_str_eq}     \\
        \rank{\ket{E(x > y)}}     &\leq k                                     \label{eq:rank_str_gt}     \\
        \rank{\ket{E(x + 1 = y)}} &\leq k+1                               \label{eq:rank_inc}
    \end{align}
\end{theorem}

Note that the equations also hold if $x$ or $y$ is a constant.
Our work permits the simulation of any gate where an upper bound on an applicable magic state can be derived through \cref{theorem:gate_rank_from_E_rank,lemma:rank_E_logical_connectives,theorem:rank_E_for_comparisons}.
We proceed to give a number of concrete examples of how these inequalities enable the derivation of other bounds.

\textbf{Application 1: MCX.} Consider the multi-controlled gate $C^k \XGate$, also known as the MCX gate.
This gate is omnipresent in quantum singular value transformations~\cite{Gilyen19Qsvt} as well as in circuits performing arithmetic~\cite{Martyn2021GrandUnificationOfQuantumAlgorithms}.
It applies an $\XGate$ when all $k$ control bits $x_1,\dots, x_k$ are $1$.
This means $C^k \XGate=C_{x_1 \dots x_k = 1^k} \XGate$.
Using \cref{theorem:gate_rank_from_E_rank,theorem:rank_E_for_comparisons}, we obtain 
\begin{align*}
    \rank{C^k \XGate} &= \rank{C^k \RxGate(\pi)}                             & \text{[$\XGate=\RxGate(\pi)$]} \\
    &=\rank{C^k \RzGate(\pi)}                                                & \text{[\cref{eq:rank_C_phi_Z_eq_rank_C_phi_X}]} \\
    &\leq \rank{\ket{E(x_1 \dots x_k = 1^k)}} + 1                                  & \text{[\cref{eq:rank_C_phi_Rz_bound_E}]} \\
    &\leq 2                                                                  & \text{[\cref{eq:rank_str_eq}]}
\end{align*}

Combining with \cref{theorem:main_algorithm}, we see that stabilizer circuits with $t$ $C^k \XGate$ gates can be strongly simulated in $O(2^t n^2 (m + k t))$ time.
In particular, observe that the exponential factor of $2^t$ is independent of $k$, as opposed to $2^{O(k t)}$ achieved by standard compilation-driven techniques~\cite{Zindorf2024efficient_implementation_multi_controlled_quantum,Vale2023decomposition_of_multi_controlled_special_unitary_single_qubit_gates}.
This recovers a result of~\cite{Koch_2024}.

\textbf{Application 2: $\pmb{C^k U}$.} 
Consider a $C^k U$ gate for an arbitrary, single-qubit unitary $U$.
As $C^k U=C_{x_1 \dots x_k = 1^k} U$, and $\rank{\ket{E(x_1 \dots x_k = 1^k)}}=1$, we apply \cref{eq:rank_C_phi_U} to obtain $\rank{C^k U}\leq 8$.
Then a circuit with $t$ $C^k U$ gates, such as a CVO-QRAM circuit~\cite{de_Veras2022_CVO_QRAM_quantum_algorithm}, suffers an exponential factor of $8^t$, again avoiding an exponential dependency on $k$.

\textbf{Application 3: An oracle gate.} 
Consider the parameterized gate $C_{x \geq y}\RzGate(\theta)$.
By combining \cref{eq:rank_str_gt} with \cref{eq:rank_E_negation}, we see that
\begin{align*}
    \rank{\ket{E(x\geq y)}}=\rank{\ket{E(\neg (y > x))}} \leq \rank{\ket{E(y > x)}} + 1 \leq k+1.
\end{align*}
By \cref{eq:rank_C_phi_Rz_bound_E} and \cref{eq:rank_str_gt}, $\rank{C_{x \geq y}\RzGate(\theta)}\leq k+2$, so by \cref{theorem:main_algorithm},
we obtain an exponential factor of $(k+2)^t=2^{O(t\log k)}$, an improvement over the compilation-driven $2^{O(kt)}$~\cite{remaud2024optimizing}.

With \cref{theorem:gate_rank_from_E_rank,theorem:rank_E_for_comparisons,lemma:rank_E_logical_connectives} concerning the rank of conditional single-qubit gates, a natural question is whether we can achieve similar performance improvements for multi-qubit gates.
We consider query gates, a standard way to provide quantum access to an arbitrary Boolean mapping $\{0,1\}^{\ell}\arrow{}\{0,1\}^{\ell}$, which are used, e.g., in Simon's algorithm~\cite{Simon1997OnThePowerOfQuantumComputation} and Shor's algorithm for period finding~\cite{Shor1994AlgorithmsForQuantumComputation}.
Our results on Boolean mappings are captured in \cref{theorem:rank_of_effectful_gate}.

\smallskip
\begin{theorem}
    \label{theorem:rank_of_effectful_gate}
    Let $\booleanMapping\colon\{0,1\}^\ell\xrightarrow[]{}\{0,1\}^{\ell}$ be a Boolean mapping, and define the query gate $U_\booleanMapping$ to compute
$$\ket{x}\ket{0^\ell} \arrow{U_\booleanMapping} \ket{x}\ket{\booleanMapping(x)}.$$
Then we have
\begin{align}
    \rank{U_\booleanMapping} &\leq \rank{\ket{E(y=\booleanMapping(x))}} + 1 \label{eq:rank_of_U_f} \\
    \rank{C_\varphi U_\booleanMapping} &\leq \rank{\ket{E(\varphi)}} (\rank{\ket{E(y=\booleanMapping(x))}}+1) + 2 \label{eq:rank_of_C_varphi_U_f}
\end{align}
Moreover, we have
\begin{align}
    \rank{\ket{E(y=\booleanMapping(x))}} &\leq 2^{\ell} \label{eq:rank_y_eq_f_of_x_generally} 
\end{align}
\end{theorem}

Although the bound of \cref{eq:rank_y_eq_f_of_x_generally} is exponential in $\ell$, note that the exponent is half the number of qubits that $U_{\booleanMapping}$ acts on (i.e., $2\ell$ qubits). 
While this is a general upper bound, tighter bounds can be obtained for specific choices of $\booleanMapping$.

\textbf{Application 4: Increment.}
One such example is the increment gate $\INCGate_\ell$, which computes 
\[\ket{x}\ket{0^\ell}\arrow{\INCGate_\ell} \ket{x}\ket{(x+1) \mod 2^\ell}\]
for $\ell$-bit numbers $x$.
This gate is a useful component in quantum convolution~\cite{Jeng2023GeneralizedQuantumConvolutionforMultidimensionalData}, solving time-dependent linear differential equations~\cite{fang2023time}, and in quantum walks~\cite{douglas2009efficient}.
Combining \cref{theorem:rank_of_effectful_gate} with \cref{eq:rank_inc} of~\cref{theorem:rank_E_for_comparisons}, we see that $\rank{\INCGate_\ell} \leq \ell+2$.
Using compilation~\cite{Gidney2015IncrementGates}, we instead obtain $\rank{\INCGate_\ell}=2^{O(\ell)}$.

As we have derived low (polynomial) bounds on the rank of certain high-level gates captured in our theorems, a natural question is whether other high-level gates also have polynomial rank.
We consider the $\ADDGate_\ell$, $\MULGate_\ell$, and $\QFTGate_\ell$ gates, implementing $\ell$-bit addition, Quantum Fourier Transform, and multiplication, 
respectively.
By applying the Choi-Jamiołkowski isomorphism~\cite{Choi1975,Jamiolkowski1972}, any $\ell$-qubit unitary $U$ can be implemented using a magic state $\frac{1}{\sqrt{2^\ell}}\sum_{x}\ket{x}(U\ket{x})$, which, by taking the basis vectors as the stabilizer decomposition, gives upper-bounds $\rank{\ADDGate_\ell}\leq 2^{2\ell}$, $\rank{\MULGate_\ell}\leq 2^{2\ell}$, and $\rank{\QFTGate_\ell}\leq 2^{2\ell}$, respectively.
The next theorem states that the ranks of these gates are indeed expected to grow exponentially in $\ell$, and thus significant improvements over the aforementioned implementations are unlikely.

\smallskip
\begin{theorem}
\label{theorem:hyperpolynomial_rank}
    Each of $\rank{\ADDGate_{\ell}}$, $\rank{\MULGate_\ell}$, and $\rank{\QFTGate_\ell}$ is hyper-polynomial unless $\PeqNP$, and is $2^{\Omega(\ell)}$ under ETH.
\end{theorem}

\cref{theorem:hyperpolynomial_rank} is derived by constructing, for each of the target gates, a family of circuits which contains a fixed number of such gates, and requires exponential time to simulate, under ETH~\cite{Impagliazzo99complexity_of_k_sat}.
Then applying \cref{theorem:main_algorithm} to the circuits, we find that the rank must be exponential under the hypothesis.
For the $\ADDGate_\ell$, $\MULGate_\ell$, and $\QFTGate_\ell$ gates, our proofs further establish that the corresponding $2^{2\ell}$ upper bound on the stabilizer rank is asymptotically tight on the exponent.
It is known that circuits composed of Clifford and arbitrarily many $\ADDGate_\ell$ gates, each acting on $\ell=O(1)$ qubits, are universal for quantum computing and are generally expected to have hyper-polynomial rank.
However, the use of arbitrarily many $\ADDGate_{\ell}$ gates is insufficient to lower bound the stabilizer rank of a single $\ADDGate_{\ell}$ gate. 
In contrast, our reduction uses circuits composed of Clifford and a constant number of $\ADDGate_\ell$ gates (and only one $C^\ell \XGate$ gate), where $\ell$ is not bounded. 
This allows us to establish lower bounds on the rank of $\ADDGate_{\ell}$ (and likewise for $\MULGate_\ell$ and $\QFTGate_\ell$).

\textbf{Practical contributions.}
We implement our simulator behind \cref{theorem:main_algorithm} and evaluate its performance on simulating various benchmarks containing high-level gates, the stabilizer rank of which we have bounded in \cref{theorem:gate_rank_from_E_rank}.
In particular, we benchmark on CVO-QRAM state preparation circuits~\cite{de_Veras2022_CVO_QRAM_quantum_algorithm}, as well as oracle gates and instances of Grover's algorithm.
In all cases, we compare the performance of our direct simulation to the performance of the Qiskit Aer simulation library~\cite{qiskit_aer_2024}, which compiles the high-level circuit before simulating it.
In all cases, our direct simulator outperforms the Qiskit Aer baselines by several orders of magnitude as instance sizes grow.

\section{Preliminaries}\label{sec:preiminaries}
Throughout this paper, we use $a,b,\dots \in \{0,1\}$ to denote single (classical) bits,
and $x,y,\dots \in\{0,1\}^n$ to denote bit-strings of length $n$, where $n$ is either explicit or clear from the context.
Given such a bit-string $x$ and some $j\in\{1,\dots, n\}$, we denote by $x_j\in \{0,1\}$ the $j$'th bit of $x$.
For two bit-strings $x$ and $y$ of equal length, we write $x \oplus y$ for their bitwise XOR operation. 

\Paragraph{Quantum states.}
An $n$-qubit quantum state is a $2^n$-dimensional vector 
$$\ket{\psi} = \sum_{x\in\{0,1\}^n} \psi_x \ket{x}$$
where the set of basis states $\{\ket{x}\}_{x\in\{0,1\}^n}$ form an orthonormal basis for a $2^n$-dimensional vector space,
and $\psi_x \in \mathbb{C}$ are complex numbers satisfying $\sum_{x\in\{0,1\}^n} |\psi_x|^2 = 1$.
For convenience, we often write states such as $\ket{0}+\ket{1}$, with the understanding that they are scaled by real, non-negative constants such that $\sum_{x\in\{0,1\}^n} |\psi_x|^2 = 1$.
To be explicit about this, we write, e.g., $\ket{\psi}\propto \ket{0}+\ket{1}$ to mean that, $\ket{\psi}= \frac{1}{\sqrt{2}}(\ket{0}+\ket{1})$.
For quantum states $\ket{\psi_1}$ and $\ket{\psi_2}$, we write their (tensor) product state as $\ket{\psi_1}\ket{\psi_2}$. 
One state that is used throughout this paper is the uniform superposition of $n$ qubits, denoted as
\[
\ket{+^n}\propto\sum_{x\in\{0,1\}^n}\ket{x}\ .
\]

\Paragraph{Quantum gates and circuits.}
A quantum state evolves by applications of quantum gates.
A quantum gate $\GATE$ corresponds to a $2^n \times 2^n$-dimensional unitary matrix $U_\GATE$.
We write $U_\GATE^{\dagger}$ for the conjugate transpose of $U_\GATE$.
Applying $\GATE$ to a state $\ket{\psi}$ means multiplying $U_\GATE$ onto $\ket{\psi}$, obtaining the state $U_\GATE \ket{\psi}$.
Since 
\[
U_\GATE \ket{\psi} = \sum_{x\in\{0,1\}^n} \psi_x U_\GATE \ket{x},
\]
a quantum gate can be described by how it acts on the basis states.
Given unitaries $U$, $U_1$, $U_2$, we write $U_1\otimes U_2$ for the tensor product between $U_1$ and $U_2$, and $U^{\otimes k}$ for the $k$-fold tensor product of $U$.
If a gate $D$ is diagonal (i.e., all non-diagonal entries of $U_D$ are $0$), we write $D_x$ for the $(x,x)$-th entry of $U_D$, with $x\in\{0,1\}^{n}$.
A diagonal gate $D$ has the effect
$
\ket{x} \arrow{D} D_x \ket{x}
$
for each $x$.

A circuit $C$ is a sequence of gates, where the $i$'th gate corresponds to a unitary $U_i$. 
Then $C$ implements the unitary $U_C=U_mU_{m-1}\cdots U_1$.
For convenience, given a state $\ket{\psi}$, we write $C\ket{\psi}$ to denote the state $U_C\ket{\psi}$.
We say that two unitaries $U_1$ and $U_2$ are equivalent (up to a global phase), written as $U_1 \equiv U_2$, if there is a real number $\theta$ such that we have $U_1 = e^{i\theta} U_2$.
For circuits $C_1$ and $C_2$, we take $C_1 \equiv C_2$ to mean $U_{C_1} \equiv U_{C_2}$, and $C_1 \equiv U_2$ to mean $U_{C_1} \equiv U_2$.

\Paragraph{Clifford gates.}
The Clifford gate set consists of the Phase ($\SGate$), Hadamard ($\HGate$), and CNOT ($\CXGate$) gates, and constitutes an important family of quantum gates, each acting on the basis states as follows.
\begin{align*}
    \ket{a} \arrow{\SGate} i^{a} \ket{a} \qquad\qquad
    \ket{a} \arrow{\HGate} \frac{1}{\sqrt{2}} \sum_{b=0}^1 (-1)^{ab}\ket{b} \qquad\qquad
    \ket{a,b} \arrow{\CXGate} \ket{a,a\oplus b}
\end{align*}
The Clifford gates suffice to construct other gates such as $\XGate$ and $\ZGate$, and can be used to construct quantum states such as an EPR pair $1/\sqrt{2} (\ket{00}+\ket{11})$.

\Paragraph{Measurements.}
While gates are invertible and affect the state deterministically, measurements are probabilistic and are not invertible.
Suppose we measure the $j$'th bit of a state $\ket{\psi}=\sum_x \psi_x \ket{x}$ in the 0,1-basis.
The probability of outcome $b\in \{0,1\}$ of the measurement is $\sum_{x\colon x_j=b} |\psi_x|^2$.
After measurement, the state is updated to $\ket{\psi'} \propto \sum_{x\colon x_j=b} \psi_x \ket{x}$.
This notion naturally generalizes to simultaneous measurements on several qubits.

In simulation, as opposed to the real world, measurement outcomes can be post-selected, meaning the outcome is given instead of being sampled\footnote{Naturally, we cannot post-select on an outcome that has $0$ probability, as this would collapse the state of the system to the zero vector which is not a quantum state.}.
As an example, given a state $1/\sqrt{2} (\ket{00}+\ket{11})$, the post-selected measurement of the first qubit to $0$ is $\ket{00}$.

\Paragraph{Stabilizer circuits.}
Together, Clifford gates and 0,1-measurements are called stabilizer gates~\cite{Gottesman1996_source_for_stabilizer_formalism}.
Circuits consisting only of stabilizer gates are called stabilizer circuits.
Any state of the form $C \ket{0^n}$ where $C$ is a stabilizer circuit is called a stabilizer state.
Concrete examples include any basis vector $\ket{x}$, for $x\in \{0,1\}^n$, as well as the uniform superposition state $\ket{+^n}$.

While stabilizer circuits can be efficiently simulated on a classical computer~\cite{Aaronson04sim_of_stabilizer_circuits}, they are not universal for quantum computing. States that cannot be expressed as $C \ket{0^n}$, for a stabilizer circuit $C$, are called non-stabilizer states, and can only be reached using some non-stabilizer gates.

\Paragraph{Universal circuits.}
A common choice of a non-stabilizer gate is the $\TGate$ gate, defined as 
$
\ket{a} \arrow{\TGate} (e^{i\pi/4})^{a} \ket{a}
$.
Another choice is the $\RzGate$ gate, a gate which is parameterized by a real number $\theta$, with
$
\ket{a} \arrow{\RzGate(\theta)} (e^{i\theta})^{a} \ket{a}
$.
Access to $\TGate$ gates or $\RzGate$ gates makes stabilizer circuits universal for quantum computation, meaning that any quantum state can be approximated to arbitrary precision~\cite{Dawson2005solovay_kitaev_thrm}.
Both $\TGate$ and $\RzGate$ are diagonal gates, hence, in order to simulate arbitrary quantum computations, it suffices to simulate stabilizer+diagonal circuits.

\Paragraph{Magic states and state injection.}
Given a circuit $C$, we can substitute any diagonal, non-stabilizer gate $D$ for a gadget which consumes an appropriate non-stabilizer state, called a \emph{magic state}~\cite{Zhou_2000_gadgetize_T_gates_magic}, defined as
\[
\ket{D}:=\sum_{x\in\{0,1\}^n} D_x \ket{x} \ .
\]
This is known as magic state injection, and is commonly simulated with post-selection~\cite{Bravyi2019simulating_q_circuits_by_low_rank_stabilizer_decomp}, which reduces the size of the state injection gadget.
Specifically, if $D$ acts on $n$ qubits, the magic state injection gadget uses $n$ $\CXGate$ gates and $n$ measurement gates with post-selection, as well as $n$ ancillary bits to store $\ket{D}$.

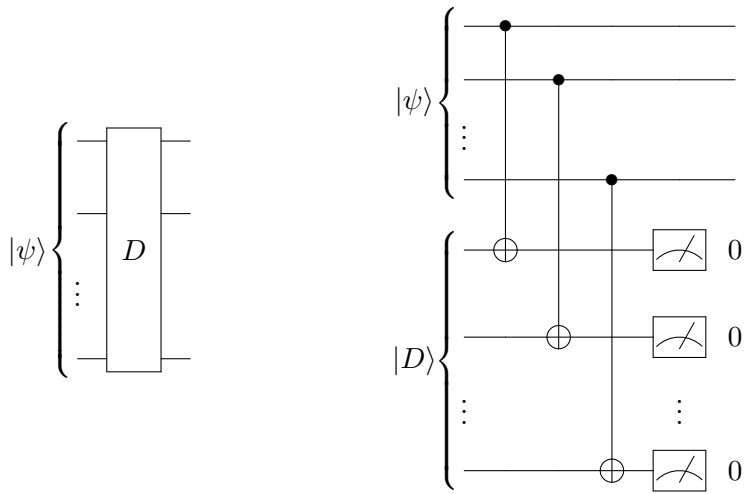
\begin{figure}[h]
\centering
\begin{subfigure}[c]{.4\textwidth}
\[
\Qcircuit @C=1em @R=1.6em {
\lstick{} & \multigate{3}{D} & \qw     \\
\lstick{} & \ghost{D}        & \qw     \\
\vdots    & \nghost{D}       &         \\
\lstick{} & \ghost{D}        & \qw     \inputgroupv{1}{4}{1.2em}{3.8em}{\ket{\psi}} \\
}
\]

\end{subfigure}
\begin{subfigure}[c]{.4\textwidth}

\[
\Qcircuit @C=1em @R=1.6em {
\lstick{} & \ctrl{4} & \qw      & \qw      & \qw    & \qw \\
\lstick{} & \qw      & \ctrl{4} & \qw      & \qw    & \qw \\
\vdots    &          &          &          &        &     \\
\lstick{} & \qw      & \qw      & \ctrl{4} & \qw    & \qw     \inputgroupv{1}{4}{1.2em}{2.6em}{\ket{\psi}} \\
\lstick{} & \targ    & \qw      & \qw      & \meter & 0  \\
\lstick{} & \qw      & \targ    & \qw      & \meter & 0  \\
\vdots    &          &          &          & \vdots &    \\
\lstick{} & \qw      & \qw      & \targ    & \meter & 0      \inputgroupv{5}{8}{1.2em}{3.8em}{\ket{D}} \\
}\]
\end{subfigure}
\caption{An implementation of a diagonal gate via magic state injection. \emph{Left:} a diagonal gate $D$.
\emph{Right:} an equivalent state injection gadget consuming a magic state $\ket{D}$.
The gadget uses state teleportation and measurements with post-selection on the outcome $00\dots 0$ to apply $D$ on $\ket{\psi}$.}
\label{fig:state_injection}
\end{figure}

\cref{fig:state_injection} provides an illustration.
The circuit performs the transformation $\ket{x} \arrow{} D_x \ket{x}$ via the following steps:
\begin{align*}
    \ket{x}\ket{D} 
        &\propto \sum_{y\in\{0,1\}^n} D_y \ket{x}\ket{y} \\
        &\arrow{\CXGate^{\otimes k}} \sum_{y\in\{0,1\}^n} D_{x\oplus y} \ket{x}\ket{y} \\
        &\arrow{Meas.} D_{x \oplus 0^k} \ket{x}\ket{0^k} 
        = D_x \ket{x}\ket{0^k}
\end{align*}
Discarding the bits storing $\ket{0^k}$, we obtain $D_x \ket{x}$ as desired.
As magic state injection gadgets consist only of stabilizer gates, the task of simulating arbitrary quantum circuits is thereby reduced to simulating stabilizer circuits on magic states. 
Although magic states are not (generally) stabilizer states, they can be represented as superpositions of stabilizer states, a process known as stabilizer decomposition.

\Paragraph{Stabilizer rank.}
Any arbitrary (not necessarily stabilizer) state can be represented as a weighted sum
\begin{align}
    \ket{\psi} = \sum_{j=1}^{k} c_j \ket{\psi_j}\ , \label{eq:def_stabilizer_rank}
\end{align}
where $\ket{\psi_j}$ are stabilizer states and $c_j$ are complex constants.
Indeed, one can always pick $k=2^n$ and let $\ket{\psi_j}$ be the basis states $\ket{0^n},\dots,\ket{1^n}$, giving the definition of a quantum state.
The smallest value of $k$ for which \cref{eq:def_stabilizer_rank} holds is known as the stabilizer rank $\rank{\ket{\psi}}$ of $\ket{\psi}$~\cite{Bravyi16trading_classical_and_q_resources}.
Note that by definition, $\rank{\ket{\psi}}=1$ iff $\ket{\psi}$ is a stabilizer state.
Bravyi et al. \cite{Bravyi16trading_classical_and_q_resources} show that the stabilizer rank is submultiplicative, i.e., for all states $\ket{\phi}$ and $\ket{\psi}$, we have
\begin{align}
    \rank{\ket{\phi} \ket{\psi}}\leq \rank{\ket{\phi}} \rank{\ket{\psi}}. \label{eq:product_state_rank_upper_bound}
\end{align}

\section{A Gadget-based Simulation Algorithm}\label{sec:simulation}

In this section, we turn our attention to the simulation of circuits containing high-level non-stabilizer gates, that are generally non-diagonal, proving \cref{theorem:main_algorithm}.
Towards this, we first generalize the concept of stabilizer rank from diagonal gates to arbitrary unitaries.
This allows us to lift the standard simulation algorithm based on state injection of diagonal gates~\cite{Bravyi2019simulating_q_circuits_by_low_rank_stabilizer_decomp} to an algorithm handling arbitrary unitaries, with its complexity depending on the stabilizer rank of such unitaries.
In later sections, we study the stabilizer rank of various high-level gates of general interest, thereby improving the simulation complexity of circuits containing such gates.

\Paragraph{Stabilizer rank of arbitrary gates.}
For a stabilizer+diagonal circuit $C$ with non-stabilizer diagonal gates $D_1, D_2, \dots, D_k$, we define the magic state of $C$ as $\ket{C}:=\ket{D_1}\ket{D_2}\dots\ket{D_k}$. 
This yields $\rank{\ket{C}}=\rank{\ket{D_1}\ket{D_2}\dots\ket{D_k}}$.
Next, we lift the notion of stabilizer rank to arbitrary unitaries.
In particular, given some arbitrary gate $\GATE$, we define the stabilizer rank of $\GATE$ as the minimum stabilizer rank of the magic state among all stabilizer+diagonal circuits which are equivalent to $\GATE$:
\begin{align}
    \UnitaryRank{\GATE}:=\min_C\{\rank{\ket{C}}\colon C\equiv \GATE \text{ and } C \text{ is a stabilizer+diagonal circuit}\} \ . \label{def:rank_arbitrary_gate}
\end{align}
Observe that if $C$ is a stabilizer+diagonal circuit with corresponding unitary $U_C$, we have $\UnitaryRank{U_C}\leq \rank{\ket{C}}$.
A strict inequality $\UnitaryRank{U_C} < \rank{\ket{C}}$ implies that there exists another stabilizer+diagonal circuit $C'$ which
(i)~defines the same unitary up to global phase, i.e., $C'\equiv C$, and
(ii)~its associated magic state has a smaller rank, i.e., $\rank{\ket{C'}}<\rank{\ket{C}}$.
For example, consider a unitary $U_C$ implemented by $C=\TGate \TGate \TGate$. 
Here $\ket{C}=\ket{\TGate}\ket{\TGate}\ket{\TGate}$, giving $\rank{\ket{C}}=3$~\cite{Gosset2021improved_upper_bounds_stabilizer_rank}.
By using the identity $\SGate = \TGate \TGate$,
we see that $U_C$ is also implemented as $C'=\SGate \TGate$.
Here we obtain $\ket{C'}=\ket{\TGate}$, giving $\rank{\ket{C'}}=2<\rank{\ket{C}}$, since the state $\ket{\TGate}\propto \ket{0}+\sqrt{i}\ket{1}$ has stabilizer rank $2$.

\Paragraph{Low-rank implementation strategies.} 
One way to simulate an arbitrary circuit is to first substitute its non-stabilizer gates for stabilizer+diagonal sub-circuits.
The choice for such a substitution, and in particular the rank of each sub-circuit can significantly impact performance.
We implement each non-stabilizer gate $\GATE$ by a stabilizer circuit $\impl{\GATE}$ applied to a magic state $\ket{\GATE}$.
In particular, $\ket{\GATE}$ is given as $(c_1, C_1),(c_2, C_2),\dots, (c_k, C_k)$ for stabilizer circuits $C_j$ and constants $c_j$, encoding $\ket{\GATE}=\sum_{j=1}^k c_j C_j\ket{00\dots 0}$.
We call this a $(\MaxGadgetSize,k)$-decomposition of $\GATE$ where naturally, we have $\rank{\GATE}\leq k$.
We require that $\impl{\GATE}$ uses $O(n)$ ancillae, and that $\impl{\GATE}$ and each of $C_1,\dots,C_k$ is over $\leq \MaxGadgetSize$ gates.
For correctness, we also require that $(\GATE\ket{\psi})\ket{00\dots0}\equiv \impl{\GATE}\ket{\psi}\ket{\GATE}$.

\Paragraph{Strong simulation.}
Given a circuit $C$ and a bit-string $x$, the task of strong simulation is to compute the probability of the state $\ket{x}$ after $C$ has been executed on $\ket{0^n}$, written $|\braket{x|C|0^n}|^2$.
If $C$ contains non-stabilizer gates $\GATE_1, \GATE_2,\dots, \GATE_t$, the above probability can be computed by
(i)~substituting each $\GATE_j$ with a $(\MaxGadgetSize_j,k_j)$-decomposition $\impl{\GATE}_j$, thereby obtaining a ``gadgetized'' stabilizer circuit $C'$ consuming a magic state $\ket{C'}=\ket{\GATE_1}\ket{\GATE_2}\dots \ket{\GATE_t}$, and
(ii)~evaluating
\[
|\bra{x}\bra{00\dots 0}C'\ket{0^n}\ket{C'}|^2=|\braket{x|C|0^n}|^2\ .\numberthis\label{eq:strong_simulation_magic_state}
\]
Using \cref{eq:product_state_rank_upper_bound}, we see that the stabilizer rank of the magic state $\ket{C'}$ is upper-bounded by
\[
\rank{\ket{C'}}\leq \prod_{j=1}^t k_j \ .
\]
Let $\chi:=\prod_{j=1}^t k_j$, and thus we can write $\ket{C'}=\sum_{j=1}^{\chi} c_j \ket{\psi_j}$, where each $c_j=a_1 a_2 \dots a_t$ is a product of unique choices of coefficients in the representations of the respective $\ket{\GATE_1}\ket{\GATE_2}\dots\ket{\GATE_t},$ and $\ket{\psi_j}$ is the tensor product of the corresponding stabilizer states (and thus a stabilizer state).
Substituting this sum into \cref{eq:strong_simulation_magic_state}, we obtain
\begin{align*}
    \Big|\bra{x}\bra{00\dots 0}C'\ket{0^n}\ket{C'}\Big|^2 
    &= \Big|\bra{x}\bra{00\dots 0}\sum_{j=1}^{\chi} c_j C' \ket{0^n}\ket{\psi_j}\Big|^2 \\
    &=\Big|\sum_{j=1}^{\chi} c_j \bra{x}\bra{00\dots 0}C' \ket{0^n}\ket{\psi_j}\Big|^2\ . \numberthis\label{eq:strong_simulation_sum}
\end{align*}
We evaluate the sum of \cref{{eq:strong_simulation_sum}} by performing $\chi$ simulations, each of the stabilizer circuit $C'$ applied to a stabilizer state $\ket{0^n}\ket{\psi_j}$, and then weighing and summing the resulting values $\bra{x}\bra{00\dots 0}C' \ket{0^n}\ket{\psi_j}$.
Strong simulation algorithms of stabilizer circuits generally provide $|\bra{x}\bra{00\dots 0}C' \ket{0^n}\ket{\psi_j}|^2$, 
which does not suffice to compute $\bra{x}\bra{00\dots 0}C' \ket{0^n}\ket{\psi_j}$, as the complex phase is lost.
Hence, this approach requires a phase-sensitive simulation.
Using directly the phase-sensitive simulator of~\cite{Bravyi2019simulating_q_circuits_by_low_rank_stabilizer_decomp} $\chi$ times, 
we obtain an algorithm with a total run-time of $O(\chi  n_g^2 m_g)$, where $n_g$ and $m_g$ are the number of qubits and number of gates of $C'$, respectively.

We can relate $n_g$ and $m_g$ to the number of qubits $n$ and gates $m$ of the original circuit $C$, as follows.
Each magic state of $\impl{\GATE}_j$ is over $O(n)$ qubits, implying that $\ket{C'}$ is a state over $O(t n)$ qubits, therefore $n_g=O(t n)$.
Moreover, each $\GATE_j$ in $C$ is replaced by a $(\MaxGadgetSize_j,k_j)$-decomposition $\impl{\GATE}_j$ in $C'$.
Defining $\MaxGadgetSize:=\max_j \MaxGadgetSize_j$, we have $m_g\leq m + 2t \MaxGadgetSize$.
Thus, the complexity of the simulation is $O(\chi t^2 n^2 (m+ t \MaxGadgetSize))$.

In the remainder of this section, we outline some simple optimizations that remove the $t^2$ factor from the above run-time, thereby arriving at \cref{theorem:main_algorithm}.
In later sections, we provide $(\MaxGadgetSize,k)$-decompositions of various natural high-level gates $\GATE$ where the stabilizer rank $k$ is small and $\MaxGadgetSize=O(n)$.

\Paragraph{Efficiency improvements.}
The time and space complexity of the above simulation process can be optimized further by storing the magic state $\ket{C'}$ more efficiently.

First, note that simply storing the full magic state $\ket{C'}$ requires $\Theta(tn)$ qubits.
However, most of these qubits are only used once, which allows for the following optimization.
Let $\ket{\psi_1}$ be the first magic state consumed by $C'$.
During the simulation process, we initially store only $\ket{\psi_1}$ in ancillary qubits.
After $\ket{\psi_1}$ has been consumed, post-selecting has reset all ancillary qubits, now storing $\ket{00\dots0}$.
Thus we use the same ancillae for storing the second magic state $\ket{\psi_2}$, etc.
Overall, our simulation now operates over circuits of $O(n)$ qubits, which improves the run-time by a factor of $t^2$ compared to the baseline simulation, thereby arriving at \cref{theorem:main_algorithm}.
As a side-note, one could permit $\omega(n)$ ancillae, which would further impact the asymptotic run-time of \cref{theorem:main_algorithm}.
We do not treat this general case in this work, as $O(n)$ ancillae suffice for the high-level gates we consider in later sections.

Second, the summation in \cref{eq:strong_simulation_sum} can also be naturally parallelized, yielding a parallel run-time of $O(\log \chi + n^2 (m+t \MaxGadgetSize))=O(n^2 (m+t \MaxGadgetSize))$ on $\chi$ threads.
Each term of the sum requires $O(n^2)$ memory to run the phase-sensitive simulator of~\cite{Bravyi2019simulating_q_circuits_by_low_rank_stabilizer_decomp}.
As we perform $\chi$ such simulations in a row, we need $O(\log \chi)$ bits to track how many simulations have been performed so far.
Hence, memory is bounded by $O(\log \chi + n^2)$, plus the space required to store two complex numbers, one for the phase on the on-going stabilizer simulation, and one for the running sum. 
Note that in the worst case, every gate $\GATE$ is non-stabilizer and has a $(\MaxGadgetSize,k)$-decomposition with $k=2^{O(n)}$, thus $\chi$ is upper-bounded by $2^{O(nm)}$. 
Hence, the memory is polynomial in $n$ and $m$ (in particular, $O(nm+n^2)$), regardless of $\chi$.

One potential downside of treating each non-stabilizer gate $\GATE_j$ independently is that we might not arrive at the best possible low-rank representation of $C$.
As an example, consider a circuit $C$ consisting of two $\TGate$ gates as $\TGate \otimes \TGate$. 
Treating each of them independently by substituting for $\impl{\TGate}$ yields $\chi=4$, since $\ket{\TGate} \propto \ket{0}+\sqrt{i}\ket{1}$.
However, the combined gate $\TGate \otimes \TGate$ has a direct low rank magic state representation as $\ket{\TGate \otimes \TGate}\propto (\ket{00}+i\ket{11})+\sqrt{i}(\ket{01}+\ket{10})$, yielding a rank of $2$~\cite{Gosset2021improved_upper_bounds_stabilizer_rank}.
As a result, the bound of \cref{theorem:main_algorithm} is not tight.

\section{Upper-bounds on the Stabilizer Rank}\label{sec:upper_bounds_stabilizer_rank}

We now turn our attention to various natural high-level gates $\GATE$, and provide bounds on their stabilizer rank $\rank{\GATE}$, towards \cref{theorem:gate_rank_from_E_rank,theorem:rank_E_for_comparisons,theorem:rank_of_effectful_gate}.
We do so by presenting $(\MaxGadgetSize,k)$-decompositions $\impl{\GATE}$ for which $k$ is small.
In all cases, $\impl{\GATE}$ uses $O(n)$ gates, i.e., $\MaxGadgetSize=O(n)$.
We will be using ancillary qubits in our constructions, but the qubit count remains $O(n)$, in particular avoiding an asymptotic impact on the run-time of \cref{theorem:main_algorithm}.

\subsection{Bounds on the Rank of Conditional Single-Qubit Gates}

Conditional single-qubit gates are gates of the form $C_\varphi U$ where $U$ is a single-qubit unitary.
We derive \cref{theorem:gate_rank_from_E_rank} by establishing upper bounds on $\rank{C_\varphi U}$ for different choices of such $C_\varphi U$.

\Paragraph{Bounds on {\normalfont$\rank{C_\varphi \RzGate}$}.}
We apply controls to $\RzGate(\theta)$ specified by a Boolean predicate $\varphi \colon \{0,1\}^k\arrow{} \{0,1\}$.
We define 
\[C_\varphi \RzGate(\theta)\ket{x,b} =
\begin{cases} 
    e^{i\theta}\ket{x,b}, & \text{if } \varphi(x) \wedge b=1 \\ 
    \ket{x,b},  & \text{otherwise}
\end{cases}
\]
We implement this via the magic state $\ket{C_\varphi \RzGate(\theta)} \propto \sum_{x,b} (e^{i\theta})^{\varphi(x) \wedge b} \ket{x,b}$.
We write this as
\begin{align*}
    \ket{C_\varphi \RzGate(\theta)} &\propto \sum_{x,b} \ket{x,b} + (e^{i\theta} - 1) \sum_{x\colon\varphi(x)} \ket{x,1}
\end{align*}
where $x$ ranges over $\{0,1\}^k$, and $b$ ranges over $\{0,1\}$.
Given a Boolean function $\varphi$, we define its ``effectual'' state $E$ as
$$\ket{E(\varphi)}\propto \sum_{x\in \{0,1\}^k\colon\varphi(x)} \ket{x},$$
where $x$ ranges over $\{0,1\}^k$.
Then 
\begin{align}
    \ket{C_\varphi \RzGate(\theta)}\propto \sum_{x, b} \ket{x,b} + (e^{i\theta} - 1) \ket{E(\varphi)} \ket{1} \label{eq:C_varphi_Rz_magic_state_decomp}
\end{align}
where $x$ ranges over $\{0,1\}^k$, and $b$ ranges over $\{0,1\}$.
As $\ket{+^{k+1}}\propto\sum_{x,b} \ket{x,b}$ is the uniform superposition and hence a stabilizer state, we obtain \cref{eq:rank_C_phi_Rz_bound_E}:
$$\rank{C_\varphi \RzGate(\theta)}\leq 1+\rank{\ket{E(\varphi)}}.$$

To further motivate our interest in $\ket{E(\varphi)}$, we rewrite \cref{eq:C_varphi_Rz_magic_state_decomp} to obtain
$$\ket{E(\varphi)}\ket{1} \propto \ket{C_\varphi \RzGate(\theta)} - \sum_{x,b} \ket{x,b},$$
implying that $\rank{\ket{E(\varphi)}} \leq 1+\rank{C_\varphi \RzGate(\theta)}$.
Combining with \cref{eq:rank_C_phi_Rz_bound_E}, any upper- or lower bound on $\rank{\ket{E(\varphi)}}$ yields a corresponding bound on $\rank{C_\varphi \RzGate(\theta)}$.

\begin{figure}[H]
    \centering
\begin{center}

\begin{align*}
\Qcircuit @C=1em @R=1.6em {
\lstick{\ket{x}} & \qw & \qw {/}        & \gate{\varphi} \qwx[1] & \qw           & \qw \\
\lstick{\ket{b}} & \qw & \gate{\HGate}  & \gate{\RzGate(\theta)}    & \gate{\HGate} & \qw \\
}
\end{align*}

\end{center}
    \caption{An implementation of a $C_\varphi \RxGate(\theta)$ using a $C_\varphi \RzGate(\theta)$ gate and two $\HGate$ gates.}
    \label{fig:C_varphi_X_equiv_Z}
\end{figure}
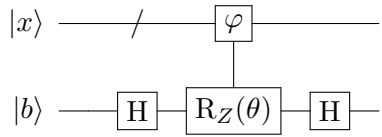

\Paragraph{Bounds on {\normalfont$\rank{C_\varphi \RxGate}$}.}
The $\XGate$ gate is a fundamental gate, providing access to the bit-flip operation.
The conditional $C_\varphi \XGate$ gate is commonly used as an oracle gate~\cite{DeutschJozsa1992RapidSolutionProblemsQuantumComputation,BernsteinVazirani1993QuantumComplexityTheory,Grover1996FastQuantumMechanicalAlgorithm,Simon1997OnThePowerOfQuantumComputation}.
A generalization of $C_\varphi \XGate$ is the $\RxGate$ gate, parameterized by a real number $\theta$, given as
$$\RxGate(\theta):=\HGate \RzGate(\theta) \HGate.$$
Specifically, we have $\XGate = \RxGate(\pi)$.
We implement $C_{\varphi}\RxGate(\theta)$ by means of $C_{\varphi}\RzGate(\theta)$, as illustrated in \cref{fig:C_varphi_X_equiv_Z}.

As $\RzGate(\theta) = \HGate \RxGate(\theta) \HGate$, by substituting the $C_\varphi \RzGate$ gate in \cref{fig:C_varphi_X_equiv_Z} for a $C_\varphi \RxGate$ gate, we obtain an implementation of a $C_\varphi \RzGate$ gate by means of the $C_\varphi \RxGate$ gate.
This gives \cref{eq:rank_C_phi_Z_eq_rank_C_phi_X}:
$$\rank{C_\varphi \RxGate(\theta)} = \rank{C_\varphi \RzGate(\theta)}.$$

\begin{figure}[H]
    \centering
\begin{center}

\begin{align*}
    \Qcircuit @C=1em @R=1.6em {
\lstick{\ket{x}} & {/} \qw & \gate{\varphi} \qwx[1] & \gate{\varphi} \qwx[1] & \gate{\varphi} \qwx[1] & \qw \\
\lstick{\ket{b}} &     \qw & \gate{\RzGate(\alpha)} & \gate{\RxGate(\beta)}  & \gate{\RzGate(\gamma)} & \qw \\
}
\end{align*}

\end{center}
    \caption{An implementation of a $C_\varphi U$ gate using two $C_\varphi \RzGate$ gates and a $C_\varphi \RxGate$.}
    \label{fig:C_varphi_U}
\end{figure}
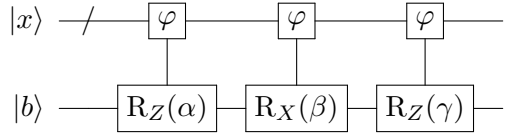

\Paragraph{Bounds on {\normalfont$\rank{C_\varphi U}$}.}
It is known that an arbitrary single-qubit unitary $U$ may be written as $$U=e^{i\delta} \RzGate(\alpha) \RxGate(\beta) \RzGate(\gamma)$$ for appropriate choices of real $\alpha$, $\beta$, $\gamma$, and $\delta$~\cite{Nielsen_Chuang2010}.
Hence, we can implement $C_\varphi U$ using two $C_\varphi \RzGate$ gates and a $C_\varphi\RxGate$ gate as in \cref{fig:C_varphi_U} to see that 
\begin{align*}
    \rank{C_\varphi U} &\leq \rank{C_\varphi \RzGate}^2\rank{C_\varphi \RxGate} & \text{[\cref{eq:product_state_rank_upper_bound}]} \\
                       &= \rank{C_\varphi \RzGate}^3 & \text{[\cref{eq:rank_C_phi_Z_eq_rank_C_phi_X}]} \\
                       &\leq (\rank{\ket{E(\varphi)}}+1)^3            & \text{[\cref{eq:rank_C_phi_Rz_bound_E}]}
\end{align*}

\begin{figure}[H]
    \centering
\begin{center}

\begin{align*}
\Qcircuit @C=0.6em @R=1.6em {
\lstick{\ket{x}} & {/} \qw & \gate{\varphi} \qwx[1] & \qw                    & \qw                   & \qw                    & \qw \\
\lstick{\ket{0}} &     \qw & \targ                  & \ctrl{1}               & \ctrl{1}              & \ctrl{1}               & \qw \\
\lstick{\ket{b}} &     \qw & \qw                    & \gate{\RzGate(\alpha)} & \gate{\RxGate(\beta)} & \gate{\RzGate(\gamma)} & \qw \\
}
\end{align*}

\end{center}
    \caption{An implementation of a $C_\varphi U$ gate using an ancillary qubit, a single $C_\varphi \XGate$ gate, two $C\RzGate$ gates, and a $C\RxGate$ gate.}
    \label{fig:C_varphi_U_alt}
\end{figure}
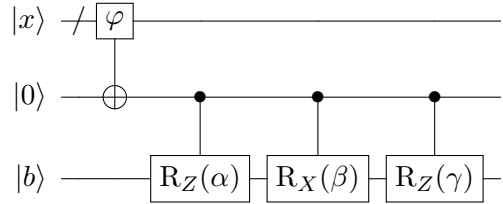

\noindent Alternatively, we can use a $C_\varphi \XGate$ gate and one ancillary qubit, as shown in \cref{fig:C_varphi_U_alt}.

Hence we also obtain 
\begin{align*}
    \rank{C_\varphi U} 
    &\leq \rank{C_\varphi \RxGate}\rank{C \RzGate}^2\rank{C \RxGate}  & \text{[\cref{eq:product_state_rank_upper_bound}]} \\
    &\leq \rank{C_\varphi \RzGate}\rank{C \RzGate}^3  & \text{[\cref{eq:rank_C_phi_Z_eq_rank_C_phi_X}]} \\
    &\leq (\rank{\ket{E(\varphi)}}+1)(\rank{\ket{E(x=1)}}+1)^3  & \text{[\cref{eq:rank_C_phi_Rz_bound_E}]} \\
    &\leq 8\rank{\ket{E(\varphi)}}+8                                           & \text{[\cref{eq:rank_str_eq}]}
\end{align*}
Then $\rank{C_\varphi U} \leq \min\{(\rank{\ket{E(\varphi)}}+1)^3,8\rank{\ket{E(\varphi)}}+8\}$.

By choosing the smaller expression based on $\rank{\ket{E(\varphi)}}$, we obtain \cref{eq:rank_C_phi_U}:
\[
\rank{C_\varphi U} \leq
            \begin{cases} 
                8,               & \text{if } \rank{\ket{E(\varphi)}}=1 \\  
                8\rank{\ket{E(\varphi)}}+8,  & \text{otherwise.}
            \end{cases}
\]

\subsection{Bounds on the Rank of Effectual States for Logical Connectives}

Predicates are often inductively defined via conjunctions, disjunctions, and negations.
In this section, we derive bounds that follow these inductive definitions.
Specifically, we prove \cref{eq:rank_E_negation,eq:rank_E_conj_bound,eq:rank_E_disj_bound} of \cref{lemma:rank_E_logical_connectives}.

\Paragraph{Bounds on $\rank{\ket{E(\varphi)}}$ for negations.}
Consider the predicate $\neg\varphi$ for some $\varphi\colon \{0,1\}^k \arrow{} \{0,1\}$.
We have
\begin{align*}
    \ket{E(\neg \varphi)} &\propto \sum_{x\in\{0,1\}^k\colon \neg \varphi(x)} \ket{x} \\
    &\propto \sum_{x\in\{0,1\}^k} \ket{x} - \sum_{x\in\{0,1\}^k\colon \varphi(x)} \ket{x} \\
    &\propto \sum_{x\in\{0,1\}^k} \ket{x} - \ket{E(\varphi)}.
\end{align*}
Hence, we obtain \cref{eq:rank_E_negation}: 
$$\rank{\ket{E(\neg \varphi)}}\leq \rank{\ket{E(\varphi)}} + 1.$$

\Paragraph{Bounds on $\rank{\ket{E(\varphi)}}$ for conjunctions.}
Consider the predicates $\varphi_1\colon \{0,1\}^{k_1} \arrow{} \{0,1\}$ and $\varphi_2\colon \{0,1\}^{k_2} \arrow{} \{0,1\}$, and let us consider the predicate $\varphi(x,y)=\varphi_1(x)\wedge \varphi_2(y)$. 
If $\varphi_1$ and $\varphi_2$ operate over disjoint inputs, here $x$ and $y$, we can write
\begin{align*}
    \ket{E(\varphi)} &\propto \sum_{x,y \colon  \varphi_1(x) \wedge \varphi_2(y)} \ket{x,y} \\
     &\propto \sum_{x \colon  \varphi_1(x)} \ket{x} \sum_{y \colon  \varphi_2(y)} \ket{y} \\
     &\propto \ket{E(\varphi_1)}\ket{E(\varphi_2)}\ .
\end{align*}
Here, $x$ ranges over $\{0,1\}^{k_1}$ and $y$ ranges over $\{0,1\}^{k_2}$.
If $\varphi_1$ and $\varphi_2$ do not operate over disjoint inputs, we use entanglement and ancillary qubits to transform the circuit to one where two predicates do operate over disjoint sets of qubits, as follows.

Suppose we want to apply a gate $C_{\varphi_1(x,y)\wedge \varphi_2(y,z)} U$ for some choice of $U$, here with $\varphi_1$ and $\varphi_2$ overlapping on the bits $y$.
Instead of applying the gate to the usual control registers $\ket{x}\ket{y}\ket{z}$, we extend with ancillary bits to obtain $\ket{x}\ket{y}\ket{00\dots0}\ket{z}$, then apply $\CXGate$ gates from the $\ket{y}$ register to the new $\ket{00\dots0}$ register to obtain $\ket{x}\ket{y}\ket{y}\ket{z}$.
We then apply a slightly modified gate $C_{\varphi_1(x,y_1)\wedge \varphi_2(y_2,z)} U$, where $y_1$ (resp. $y_2$) denotes the first (resp. second) register equal to $\ket{y}$.
Finally, we reapply $\CXGate$ gates to reset the second register storing $\ket{y}$ to $\ket{00\dots0}$, allowing it to be reused.
Since we already use $O(n)$ qubits in our simulation, these ancillaries are subsumed in the big-O notation.

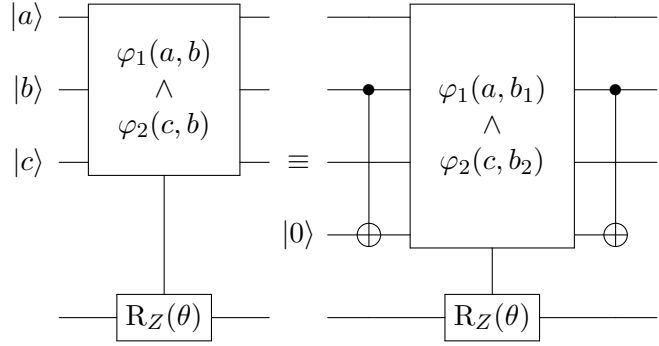
\begin{figure}[H]
    \centering
\begin{center}

\newcommand{\CONJ}{\begin{array}{c}\varphi_1(a,b) \\ \wedge \\ \varphi_2(c,b)\end{array}}
\newcommand{\CONJJ}{\begin{array}{c}\varphi_1(a,b_1) \\ \wedge \\ \varphi_2(c,b_2)\end{array}}

\begin{equation*}
\Qcircuit @C=1em @R=1.6em {
   \lstick{\ket{a}} & \ghost{\CONJ}                 & \qw &           &     & \qw      & \ghost{\CONJJ}                 & \qw      & \qw \\
   \lstick{\ket{b}} & \ghost{\CONJ}                 & \qw &           &     & \ctrl{2} & \ghost{\CONJJ}                 & \ctrl{2} & \qw \\
   \lstick{\ket{c}} & \multigate{-2}{\CONJ} \qwx[2] & \qw & \equiv    &     & \qw      & \ghost{\CONJJ}                 & \qw      & \qw \\
                    &                               &     & {\ket{0}} &     & \targ    & \multigate{-3}{\CONJJ} \qwx[1] & \targ    &     \\
                    & \gate{\RzGate(\theta)}        & \qw &           &     & \qw      & \gate{\RzGate(\theta)}         & \qw      & \qw \\
}
\end{equation*}

\end{center}
    \caption{An example of our rewriting of conjunctions.
    By introducing ancillae, we split functions so that conjunctions are over separate variables.}
    \label{fig:split_conjunctions_ancillary_qubit}
\end{figure}

\cref{fig:split_conjunctions_ancillary_qubit} captures an example of this transformation applied to a $C_{\varphi_1(a,b)\wedge\varphi_2(c,b)} \RzGate(\theta)$ gate, here overlapping on the bit $b$.

By applying this technique as necessary, we assume without loss of generality that conjunctions of predicates always operate over disjoint qubits.
This means in particular that 
\begin{align*}
    \rank{\ket{E(\varphi_1 \wedge \varphi_2)}} &= \rank{\ket{E(\varphi_1)}\ket{E(\varphi_2)}} \\
    &\leq \rank{\ket{E(\varphi_1)}}\rank{\ket{E(\varphi_2)}}. & \text{[\cref{eq:product_state_rank_upper_bound}]}\ ,
\end{align*}
yielding \cref{eq:rank_E_conj_bound}:
\[
\rank{\ket{E(\varphi_1 \wedge \varphi_2)}} \leq \rank{\ket{E(\varphi_1)}}\rank{\ket{E(\varphi_2)}}\ .
\]

\Paragraph{Bounds on $\rank{\ket{E(\varphi)}}$ for disjunctions.}
We now turn to the case of $\varphi_1 \vee \varphi_2$.
Using our strategy as before, we assume without loss of generality that $\varphi_1$ and $\varphi_2$ operate over disjoint sets of qubits.
We write
\begin{align*}
    \ket{E(\varphi_1 \vee \varphi_2)} &\propto \sum_{x,y : \varphi_1(x) \vee \varphi_2(y)} \ket{x}\ket{y} \\
       &\propto \sum_{x,y \colon \varphi_1(x)} \ket{x}\ket{y} + \sum_{x,y : \varphi_2(y)} \ket{x}\ket{y} - \sum_{x,y \colon \varphi_1(x) \wedge \varphi_2(y)} \ket{x}\ket{y} \\
       &\propto \ket{E(\varphi_1)}\ket{+\dots+} + \ket{+\dots+}\ket{E(\varphi_2)} - \ket{E(\varphi_1 \wedge \varphi_2)}.
\end{align*}
Here, $x$ ranges over $\{0,1\}^{k_1}$, while $y$ ranges over $\{0,1\}^{k_2}$.
Since $\ket{+}=\HGate\ket{0}$ is a stabilizer state, we have $$\rank{\ket{E(\varphi)}\ket{+\dots+}}=\rank{\ket{+\dots+}\ket{E(\varphi)}}=\rank{\ket{E(\varphi)}}.$$
Hence, we obtain \cref{eq:rank_E_disj_bound}:
\begin{align*}
    \rank{\ket{E(\varphi_1 \vee \varphi_2)}} \leq \rank{\ket{E(\varphi_1)}} + \rank{\ket{E(\varphi_2)}}+\rank{\ket{E(\varphi_1\wedge\varphi_2)}}. 
\end{align*}

\subsection{Bounds on the Rank of Effectual States for Predicates}

In this section, we turn to establishing $\rank{\ket{E(\varphi)}}$ for various choices of predicates.
Specifically, we prove \cref{eq:rank_str_eq,eq:rank_str_gt,eq:rank_inc} of \cref{theorem:rank_E_for_comparisons}.

\Paragraph{Bounds on $\rank{\ket{E(\varphi)}}$ for equality.}
We first consider functions performing comparison of binary integers $x$ and $y$ of length $k$.
As a warm-up, let us examine the case $\varphi(x,y) = (x = y)$.
We see that 
$$\ket{E(x=y)} \propto\sum_{w\in \{0,1\}^k} \ket{w}\ket{w}.$$
This state can be initialized from $\ket{0^k}\ket{0^k}$ by applying $\HGate$ to each of the first $k$ qubits, then applying $\CXGate$ with control bit $j$ and target bit $j+k$ for $1\leq j\leq k$.
Hence $\ket{E(x=y)}$ is a stabilizer state, and we have \cref{eq:rank_str_eq}: 
$$\rank{\ket{E(x=y)}}=1.$$

A special case of this result is when $y$ is a fixed basis state, which enables checking if $x$ equals a constant.

\Paragraph{Bounds on $\rank{\ket{E(\varphi)}}$ for string comparison.}
Let us now examine the function $\varphi(x,y)= (x > y)$, where $x$ and $y$ are interpreted as binary representations of natural numbers.
By definition, $x > y$ if there is a string $w$ s.t. we have $x=w1x'$ and $y=w0y'$, where $x'$ and $y'$ have length based on $w$.
Since the length of $w$ ranges from $0$ to $k-1$, we have
\begin{align*}
    \ket{E(x > y)} &\propto \sum_{x,y\in\{0,1\}^k\colon x > y} \ket{x,y} \\
    &\propto \sum_{\ell=0}^{k-1} \sum_{w\in\{0,1\}^\ell} \ket{w1+\dots+}\ket{w0+\dots+}.
\end{align*}
It is easy to see that 
$$\sum_{w\in\{0,1\}^\ell} \ket{w1+\dots+}\ket{w0+\dots+}$$
is a stabilizer state: we can initialize it from $\ket{0^k}\ket{0^k}$ by applying $\HGate$ to the first $\ell$ bits, then applying $\CXGate$ with control $j$ and target $j+k$ for $1\leq j \leq \ell$. 
Finally, applying $\XGate$ to the bit equaling $\ket{1}$, and $\HGate$ to each bit equaling $\ket{+}$, we obtain the desired state. 
Hence we write $\ket{E(x>y)}$ as a sum of $k$ stabilizer states, giving \cref{eq:rank_str_gt}:
$$\rank{\ket{E(x > y)}}\leq k.$$

\Paragraph{Bounds on $\rank{\ket{E(\varphi)}}$ for incrementing.}
We now turn to $\varphi(x,y)=(y=x+1 )$. 
Here, the strings $x$ and $y$ are understood as binary numbers, and $+$ is binary addition modulo $2^k$.
By definition, we must either have some $w$ of length $\ell$ s.t. $x=w01^{k-\ell-1}$ and $y=w10^{k-\ell-1}$, or we have exactly $x=1\dots1$ and $y=0\dots0$.
As before, we have
\begin{align*}
    \ket{E(y=x+1)} \propto& \sum_{x,y\in\{0,1\}^k\colon y=x+1} \ket{x,y} \\
    \propto&\sum_{\ell=0}^{k-1} \sum_{w\in\{0,1\}^\ell} \ket{w01\dots1}\ket{w10\dots0} \\
    &+\ket{1\dots1}\ket{0\dots0}.
\end{align*}
We can initialize 
$\sum_{w\in\{0,1\}^\ell} \ket{w01\dots1}\ket{w10\dots0}$ 
from $\ket{0^k}\ket{0^k}$ by applying $\HGate$ to the first $\ell$ bits, then applying $\CXGate$ with control $j$ and target $j+k$ for $1\leq j \leq \ell$. 
Finally, applying $\XGate$ to the bits equaling $\ket{1}$, we obtain the desired state. 
Hence it is a stabilizer state, and we obtain \cref{eq:rank_inc}:  
$$\rank{\ket{E(y=x+1)}}\leq k+1.$$

\subsection{The Stabilizer Rank of Query Gates}

\newcommand{\condBoolMapping}{h} 

Query gates are a standard technique for providing quantum access to a Boolean mapping $\booleanMapping\colon \{0,1\}^{\ell}\arrow{}\{0,1\}^{\ell}$~\cite{Shor1994AlgorithmsForQuantumComputation,Simon1997OnThePowerOfQuantumComputation}. 
Given $\booleanMapping$, we define the query gate $U_\booleanMapping$ to implement the transformation 
$$\ket{x}\ket{0^\ell} \arrow{U_\booleanMapping} \ket{x}\ket{\booleanMapping(x)}.$$

The gates captured by \cref{theorem:gate_rank_from_E_rank} only modify one of the qubits to which they are applied, whereas $U_\booleanMapping$ modifies $\ell$ qubits. 
To implement $U_\booleanMapping$, we use post-selecting measurements to affect multiple qubits. 
Through this, we prove equations \cref{eq:rank_of_U_f,eq:rank_y_eq_f_of_x_generally,eq:rank_of_C_varphi_U_f} of \cref{theorem:rank_of_effectful_gate}.
Our construction implementing $U_\booleanMapping$ is captured in \cref{fig:U_f_construction}:

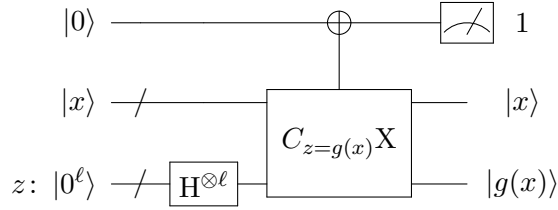
\begin{figure}[H]
    \centering
\begin{center}

\newcommand{\GN}{C_{z=\booleanMapping(x)} \XGate}

\begin{align*}
\Qcircuit @C=1em @R=1.6em {
   \lstick{\ket{0}}        &     \qw & \qw                          & \targ{}              & \meter & {1}      \\
   \lstick{\ket{x}}        & {/} \qw & \qw                          & \ghost{\GN} \qwx[-1] & \qw    & {\ket{x}} \\
   \lstick{z\colon\ket{0^\ell}}   & {/} \qw & \gate{\HGate^{\otimes\ell}}           & \multigate{-1}{\GN}  & \qw    & {\ket{g(x)}} \\
}
\end{align*}

\end{center}
    \caption{A circuit implementing $U_\booleanMapping$ by applying post-selection.}
    \label{fig:U_f_construction}
\end{figure}

The circuit performs the following transformation:
\begin{align*}
    \ket{0}\ket{x}\ket{0^\ell} &\arrow{\HGate^{\otimes \ell}} \sum_{z\in \{0,1\}^\ell} \ket{0}\ket{x}\ket{z} \\
    &\arrow{C_{z=\booleanMapping(x)} \XGate} \sum_{z\in \{0,1\}^\ell} \ket{1_{z=\booleanMapping(x)}}\ket{x}\ket{z} \\
    &\arrow{Meas.} \sum_{z\in \{0,1\}^\ell \colon z=\booleanMapping(x)} \ket{1}\ket{x}\ket{z}\\
    &= \ket{1}\ket{x}\ket{\booleanMapping(x)} 
\end{align*}

The qubit storing $\ket{1}$ can be set to $\ket{0}$ by applying an $\XGate$ gate, and can then be recycled.
In particular, we suffer no asymptotic cost in ancillae.
Doing so, we obtain the state $\ket{x}\ket{\booleanMapping(x)}$.
Using this construction, we derive \cref{eq:rank_of_U_f}: 
$$\rank{U_\booleanMapping}\leq \rank{C_{z=\booleanMapping(x)} \XGate} \leq \rank{\ket{E(y=\booleanMapping(x))}}+1.$$

Since by \cref{eq:rank_inc} of \cref{theorem:rank_E_for_comparisons}, we have $\rank{\ket{E(y=x+1)}}\leq \ell+1$, so we have the means to implement $\INCGate_\ell$:
$$\ket{x}\ket{0^\ell}\arrow{\INCGate_\ell}\ket{x}\ket{(x+1) \mod 2^\ell}.$$
Specifically, we obtain 
$$\rank{\INCGate_\ell}\leq \rank{C_{y=x+1}\XGate} \leq \ell+2.$$
As a consequence, a circuit containing a constant number of $\INCGate_\ell$ gates can be simulated in polynomial time.
Given the linear bound on incrementing, it is natural to ask if this technique can be extended to also efficiently implement other arithmetic operations, such as addition and multiplication.  
In \cref{sec:lower_bounds_stabilizer_rank}, we answer this in the negative, assuming common complexity theoretical hypotheses.

One natural question is whether this technique can be extended to general $\booleanMapping$.
From \cref{eq:rank_of_U_f}, we simply need to establish $\rank{\ket{E(y=\booleanMapping(x))}}$.
By examining 
$$\ket{E(y=\booleanMapping(x))}\propto\sum_{x\in\{0,1\}^{\ell}} \ket{x}\ket{\booleanMapping(x)},$$
we see that $\ket{E(y=\booleanMapping(x))}$ is a sum of $2^\ell$ stabilizer states of the form $\ket{x}\ket{\booleanMapping(x)}$.
This gives the upper bound of \cref{eq:rank_y_eq_f_of_x_generally}:
$$\rank{\Ket{E(y=\booleanMapping(x))}}\leq 2^\ell.$$

Another natural question is whether we can apply controls to these gates, i.e. whether we can construct gates such as $C_\varphi U_\booleanMapping$.
Specifically, we have
\[C_\varphi U_\booleanMapping \ket{x}\ket{0^\ell} =
\begin{cases} 
    \ket{x}\ket{\booleanMapping(x)}, & \text{if } \varphi(x) \\ 
    \ket{x}\ket{0^\ell},            & \text{otherwise}
\end{cases}
\]
By defining a new mapping $\condBoolMapping\colon\{0,1\}^\ell\arrow{}\{0,1\}^\ell$ as
\[\condBoolMapping(x) :=
\begin{cases} 
    \booleanMapping(x), & \text{if } \varphi(x) \\ 
    0^{\ell},  & \text{otherwise}
\end{cases}
\] 
we see that $C_\varphi U_\booleanMapping = U_\condBoolMapping$.
Hence, we straightforwardly apply \cref{eq:rank_y_eq_f_of_x_generally} to obtain a bound of $\rank{C_\varphi U_\booleanMapping}=\rank{U_\condBoolMapping}\leq 2^{\ell}+1$.
However, the construction also enables us to derive a more refined upper bound.

By \cref{eq:rank_of_U_f}, we have 
$\rank{U_\condBoolMapping}\leq \rank{\ket{E(y=\condBoolMapping(x))}}+1$,
while 
$y=\condBoolMapping(x)$ 
is equivalent to
$(\varphi(x)\wedge y=\booleanMapping(x))\vee(\neg \varphi(x) \wedge y=0^\ell)$.
Hence 
$$\rank{\ket{E(y=\condBoolMapping(x))}}=\rank{\ket{E((\varphi(x)\wedge y=\booleanMapping(x))\vee(\neg \varphi(x) \wedge y=0^\ell))}}.$$
Towards deriving a bound on this value, we see that
\begin{align*}
    \rank{\ket{E(\neg \varphi(x) \wedge y=0^\ell)}} &\leq \rank{\ket{E(\neg \varphi(x))}}\rank{\ket{E(y=0^\ell)}} & \text{[\cref{eq:rank_E_conj_bound}]} \\
    &= \rank{\ket{E(\neg \varphi(x))}}                                                            & \text{[\cref{eq:rank_str_eq}]} \\
    &\leq \rank{\ket{E(\varphi(x)}} + 1.                                                          & \text{[\cref{eq:rank_E_negation}]}
\end{align*}
Furthermore, note that
$$(\varphi(x)\wedge y=\booleanMapping(x))\wedge(\neg \varphi(x) \wedge y=0^\ell)=\bot.$$
We have
\begin{align*}
    \rank{\ket{E(y=\condBoolMapping(x))}} &=\rank{\ket{E((\varphi(x)\wedge y=\booleanMapping(x))\vee(\neg \varphi(x) \wedge y=0^\ell))}} \\
    &\leq \rank{\ket{E(\varphi(x)\wedge y=\booleanMapping(x))}} + \rank{\ket{E(\neg \varphi(x) \wedge y=0^\ell)}} \\
    & \quad + \rank{\ket{E((\varphi(x)\wedge y=\booleanMapping(x))\wedge(\neg \varphi(x) \wedge y=0^\ell))}}  & \text{[\cref{eq:rank_E_disj_bound}]} \\
    &\leq \rank{\ket{E(\varphi(x) \wedge y=\booleanMapping(x))}} + \rank{\ket{E(\varphi(x))}} + 1 + \rank{\ket{E(\bot)}} \\
    &= \rank{\ket{E(\varphi(x))}}\rank{\ket{E(y=\booleanMapping(x))}} + \rank{\ket{E(\varphi(x))}} + 1 & \text{[\cref{eq:rank_E_conj_bound}]} \\
    &= \rank{\ket{E(\varphi(x))}} (\rank{\ket{E(y=\booleanMapping(x))}}+1) + 1 
\end{align*}
Hence we obtain \cref{eq:rank_of_C_varphi_U_f}:
\[
\rank{C_\varphi U_\booleanMapping} \leq \rank{\ket{E(\varphi)}} (\rank{\ket{E(y=\booleanMapping(x))}}+1) + 2\ .
\]

This concludes our derivation of \cref{theorem:rank_of_effectful_gate}.

\section{Lower-bounds on the Stabilizer Rank}\label{sec:lower_bounds_stabilizer_rank}

We now turn our attention to other high-level gates, considering $\ADDGate_\ell$, $\MULGate_\ell$, and $\QFTGate_\ell$, which implement $\ell$-bit addition, multiplication, and Quantum Fourier Transform, respectively.
We show~\cref{theorem:hyperpolynomial_rank}, which states that the ranks of these gates cannot be polynomial assuming $\PneqNP$, while under the Exponential Time Hypothesis (ETH)~\cite{Impagliazzo99complexity_of_k_sat}, the exponent has to be linear in $\ell$ (e.g., it cannot be of the form $2^{O(\sqrt{\ell})}$).

Our proofs follow a sequence of reductions from 3-SAT, the famous $\NP$-complete problem that is defined as follows.
Given a propositional formula 
$$\varphi(x_1,\dots,x_k) := (a^1_1\vee a^1_2\vee a^1_3) \wedge \dots \wedge (a^\ell_1 \vee a^\ell_2 \vee a^\ell_3)$$ 
where $a^i_j\in \{x_p,\bar{x}_p\}_{p=1,\dots,k}$, do there exist $x_1,\dots,x_k\in \{0,1\}$ s.t. $\varphi(x_1,\dots,x_k)=1$?
We say that $\varphi$ has $k$ variables and $\ell$ clauses.
The Exponential Time Hypothesis (ETH) implies that for any algorithm solving 3-SAT, there are 3-SAT instances over $k$ variables which require $2^{\Omega(k)}$ time. 
As each $x_j$ must occur at least once, we have $k\leq \ell/3 =O(\ell)$, meaning 3-SAT has complexity $2^{\Omega(\ell)}$ under ETH.

Our sequence of reductions proceeds as follows.
First, we show that a particular high-level oracle gate, which directly computes $\varphi$, has stabilizer rank $2^{\Omega(\ell)}$ under ETH.
Second, we implement this oracle gate using two gates that perform $\ell$ parallel 3-bit logical OR operations.
Third, we implement the parallel OR gate using a symmetrically defined parallel AND gate on $\ell$ qubits.
Fourth, we implement the parallel AND gate using an $\ADDGate_\ell$ gate.
We then the parallel AND gate using a $\MULGate_\ell$ gate.
Finally, we implement an $\ADDGate_\ell$ gate using three $\QFTGate_\ell$ gates.

\begin{figure}[H]
    \centering
\begin{center}

\[
\Qcircuit @C=1em @R=1.6em {
\lstick{\ket{00\dots0}} & \qw {/}   & \gate{\HGate^{\otimes k}} & \gate{\varphi} \qwx[1]  & \gate{\HGate^{\otimes k}} & \qw      \\
\lstick{\ket{0}}        & \qw       & \qw                       & \targ{}                 & \qw                       & \qw 
}\]

\end{center}
    \caption{A circuit computing whether $\varphi$ is satisfiable.}
    \label{fig:family_hard_C_varphi_X}
\end{figure}
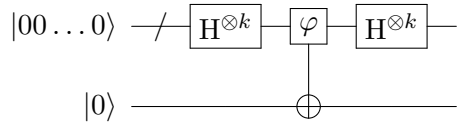

\Paragraph{Oracle gate.}
Given a 3-SAT formula  $\varphi$ over $k$ variables and $\ell$ clauses, we can compute whether it is satisfiable by simulating a single $C_\varphi \XGate$ oracle gate on a superposition of all assignments to the input variables of $\varphi$, as shown in \cref{fig:family_hard_C_varphi_X}.

The output state is $\ket{00\dots 0}\ket{0}$ iff $\varphi$ is unsatisfiable, which can be decided by simulating the circuit with target state $\ket{00\dots 0}\ket{0}$.
The circuit has one non-stabilizer gate as well as $2k=O(\ell)$ stabilizer gates and operates over $k+1=O(\ell)$ qubits.
Thus, \cref{theorem:main_algorithm} implies that the running time for its simulation is bounded by
\begin{align*}
    O&(\chi O(\ell)^2 (O(\ell)+O(\ell) 1)) = O(\rank{C_\varphi \XGate} \ell^{3})\ .
\end{align*}
This implies that $\rank{C_\varphi \XGate}$ is not polynomial under $\PneqNP$,
and $\rank{C_\varphi \XGate}=2^{\Omega(\ell)}$ under ETH.

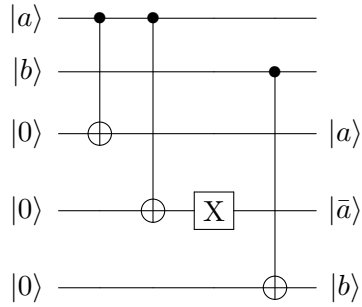
\begin{figure}[H]
    \centering
\begin{center}

\[
\Qcircuit @C=1em @R=1.6em {
\lstick{\ket{a}} & \ctrl{2} & \ctrl{3} & \qw      & \qw      & \qw  \\
\lstick{\ket{b}} & \qw      & \qw      & \qw      & \ctrl{3} & \qw  \\
\lstick{\ket{0}} & \targ{}  & \qw      & \qw      & \qw      & \qw & {\ket{a}} \\
\lstick{\ket{0}} & \qw      & \targ{}  & \gate{\XGate} & \qw      & \qw & {\ket{\bar{a}}}  \\
\lstick{\ket{0}} & \qw      & \qw      & \qw      & \targ{}  & \qw & {\ket{b}}  \\
}\]

\end{center}
    \caption{An $\INIGate$ gadget performing the transformation $\ket{ab}\ket{0^3}\arrow{\INIGate}\ket{ab}\ket{a\bar{a}b}$.}
    \label{fig:INI_gadget_example}
\end{figure}

\Paragraph{$\ORGate_\ell$ gates.}
We now proceed with implementing the oracle gate $C_\varphi \XGate$ using two $\ORGate_\ell$ gates. 
These compute $\ell$ disjunctions over three variables each, in a parallel manner:
\begin{align*}
    \ket{x}\ket{0^\ell} &\arrow{\ORGate_\ell} \ket{x} \ket{x_1\vee x_2 \vee x_3} \dots \ket{x_{3\ell-2}\vee x_{3\ell-1} \vee x_{3\ell}}\ .
\end{align*}
The following construction implies a bound on $\rank{\ORGate_\ell}$ in terms of $\rank{C_\varphi \XGate}$.
Recall that $a^i_j$ refers to the $i$-th literal of the $j$-th clause of $\varphi$.
First we introduce the $\INIGate$ gadget that performs the following transformation:
$$\ket{x}\ket{0^{3\ell}} \arrow{\INIGate} \ket{x}\ket{a^1_1 a^1_2 a^1_3} \dots \ket{a^\ell_1 a^\ell_2 a^\ell_3}\ .$$
This construction can be implemented using $\CXGate$ and $\XGate$ gates.
\cref{fig:INI_gadget_example} shows an example for the transformation $\ket{ab}\ket{0^3}\arrow{\INIGate}\ket{ab}\ket{a\bar{a}b}$.

\begin{figure}[H]
    \centering
\begin{center}

\[
\Qcircuit @C=1em @R=1.6em {
\lstick{\ket{x}}  & \qw {/} & \gate{\INIGate} & \multigate{3}{\ORGate_\ell} & \qw      & \multigate{3}{\ORGate_\ell^\dagger} & \gate{\INIGate^\dagger} & \qw \\
\lstick{\ket{0}}        & \qw     & \qw             & \ghost{\ORGate_\ell}        & \ctrl{2} & \ghost{\ORGate_\ell^\dagger}        & \qw                     & \qw \\
                        & \vdots   &                 & \nghost{\ORGate_\ell}       &          & \nghost{\ORGate_\ell^\dagger}       &                         &     \\
\lstick{\ket{0}}        & \qw     & \qw             & \ghost{\ORGate_\ell}        & \ctrl{1} & \ghost{\ORGate_\ell^\dagger}        & \qw                     & \qw \\
\lstick{\ket{0}}        & \qw     & \qw             & \qw                         & \targ{}  & \qw                                 & \qw                     & \qw \\
}\]

\end{center}
    \caption{A circuit implementing a $C_\varphi \XGate$ gate using two $\ORGate_\ell$ gates and a $C^\ell \XGate$ gate.}
    \label{fig:family_hard_OR_gate}
\end{figure}
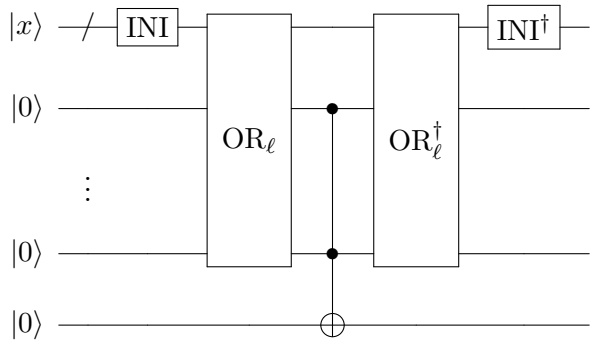

We implement the $C_\varphi \XGate$ gate via the circuit of \cref{fig:family_hard_OR_gate}, which uses the $\INIGate$ gadget twice.

As the construction implements a $C_\varphi \XGate$ gate, its magic state rank is an upper bound on the magic state rank of $C_\varphi \XGate$, i.e.,
\[
\rank{C_\varphi X}\leq \rank{C^\ell \XGate}\rank{\ORGate_\ell}^2 \leq 2\rank{\ORGate_\ell}^2\ ,
\numberthis\label{eq:ORGate_rank}
\]

where $\rank{C^\ell \XGate}\leq 2$ follows from \cref{eq:rank_C_phi_Z_eq_rank_C_phi_X,eq:rank_C_phi_Rz_bound_E} of \cref{theorem:gate_rank_from_E_rank} and \cref{eq:rank_str_eq} of \cref{theorem:rank_E_for_comparisons}.
We thus obtain that $\rank{\ORGate_\ell}$ is not polynomial under $\PneqNP$, and $\rank{\ORGate_\ell}=2^{\Omega(\ell)}$ under ETH.

\Paragraph{$\ANDGate_\ell$ gates.}
As a second intermediate step, we show that an $\ORGate_\ell$ gate can be implemented using an $\ANDGate_\ell$ gate, defined in the following manner:
\begin{align*}
    \ket{x}\ket{0^\ell} &\arrow{\ANDGate_\ell } \ket{x} \ket{x_1\wedge x_2 \wedge x_3} \dots \ket{x_{3\ell-2}\wedge x_{3\ell-1} \wedge x_{3\ell}}
\end{align*}

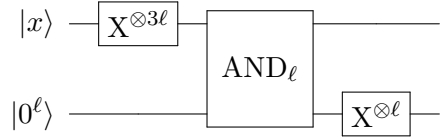
\begin{figure}[H]
    \centering
\begin{center}

\[
\Qcircuit @C=1em @R=1.6em {
\lstick{\ket{x}}      & \gate{\XGate^{\otimes 3\ell}}  & \multigate{1}{\ANDGate_\ell} & \qw                     & \qw &       \\
\lstick{\ket{0^\ell}} & \qw                       & \ghost{\ANDGate_\ell}        & \gate{\XGate^{\otimes \ell}} & \qw        \\
}\]

\end{center}
    \caption{A circuit implementing an $\ORGate_\ell$ gate using an $\ANDGate_\ell$ gate and $4\ell$ $\XGate$ gates.}
    \label{fig:family_hard_AND_gate}
\end{figure}

The construction, captured in \cref{fig:family_hard_AND_gate}, is a straightforward application of De Morgan's law, and shows that $\rank{\ORGate_\ell}\leq\rank{\ANDGate_\ell}$.
As a sidenote, if the $\ANDGate_\ell$ gate in \cref{fig:family_hard_AND_gate} is substituted for an $\ORGate_\ell$ gate, the construction instead yields an implementation of an $\ANDGate_\ell$ gate.
This shows the stronger statement that $\rank{\ORGate_\ell}=\rank{\ANDGate_\ell}$.
Combined with \cref{eq:ORGate_rank}, we obtain that $\rank{\ANDGate_\ell}$ is not polynomial under $\PneqNP$, and $\rank{\ANDGate_\ell}=2^{\Omega(\ell)}$ under ETH.

\Paragraph{$\ADDGate_\ell$ gates.}
We are now ready to derive a bound on the rank of the $\ADDGate_\ell$ gate, which implements addition over $\ell$-bit integers:
\begin{align*}
    \ket{u}\ket{v}\ket{0^\ell} &\arrow{\ADDGate_\ell} \ket{u} \ket{v} \ket{(u+v) \mod 2^\ell}.
\end{align*}
We show that an $\ANDGate_\ell$ gate can be implemented using an $\ADDGate_{3\ell}$ gate.
We do this by arranging the input bit string $x$ of $\ANDGate_\ell$ into two bit strings $u$ and $v$, padded with zeros, which will serve as the input to the $\ADDGate_{3\ell}$ gate.
In particular,
\begin{align*}
    u &:= 0&&x_1&&x_2&&0&&x_4&&x_5&&\dots&&0&&x_{\ell-2}&&x_{\ell-1} \\
    v &:= 0&&  0&&x_3&&0&&  0&&x_6&&\dots&&0&&  0&&x_{\ell}\ .
\end{align*}
Let $z=u+v$, and $z_1 z_2 \dots z_\ell$ be the bits of $z$, with $z_1$ being the most significant bit.
Then we have $z_1=x_1\wedge x_2\wedge x_3$, and generally, $z_{3j+1} = x_{3j+1}\wedge x_{3j+2} \wedge x_{3j+3}$.
Thus the bit string $z_{1}z_{4}\dots z_{3\ell-2}$ encodes the output of the $\ANDGate_\ell$ gate.
This implies that $\rank{\ANDGate_\ell} \leq \rank{\ADDGate_{3\ell}}$, implying that $\rank{\ADDGate_\ell}$ is not polynomial under $\PneqNP$, and $\rank{\ADDGate_\ell}=2^{\Omega(\ell)}$ under ETH.

\Paragraph{$\MULGate_\ell$ gates.}
We now derive a bound on the rank of the $\MULGate_\ell$ gate, which implements multiplication over $\ell$-bit integers:
$$\ket{u}\ket{v}\ket{0^\ell} \arrow{\MULGate_\ell} \ket{u}\ket{v}\ket{(uv)  \mod 2^\ell}.$$
We show that an $\ANDGate_\ell$ gate can be implemented via a $\MULGate_{7\ell}$ gate, by a process similar to the one above.
We arrange the input bit string $x$ of $\ANDGate_\ell$ in a specific pattern padded with zeros, which will serve as the input to the $\MULGate_{7\ell}$ gate.
In particular,
\begin{align*}
    u&:=000x_10x_2x_3&&000x_40x_5x_6&&\dots&&000x_{\ell-2}0x_{\ell-1} x_\ell
\end{align*}
and $v:=00\dots01001$.
Then the gate computes $uv=9u=u+8u$.
Writing out these numbers, we see the pattern
\begin{align*}
    u  &=  &&   &&   &&  0&&  0&&  0&&x_1&&  0&&x_2&&x_3&&\dots \\
    8u &= 0&&  0&&  0&&x_1&&  0&&x_2&&x_3&&  0&&  0&&  0&&\dots
\end{align*}
Then define $z=uv$, and let $z_1 z_2 \dots z_\ell$ be the bits of $z$.
Then we have $z_5=x_1\wedge x_2\wedge x_3$, while generally, $z_{7j+5} = x_{3j+1}\wedge x_{3j+2} \wedge x_{3j+3}$.
Thus the bit-string $z_5 z_{12} \dots z_{7\ell-2}$ encodes the output of the $\ANDGate_\ell$ gate.
This implies that $\rank{\ANDGate_\ell}\leq \rank{\MULGate_{7\ell}}$, implying that $\rank{\MULGate_\ell}$ is not polynomial under $\PneqNP$, and $\rank{\MULGate_\ell}=2^{\Omega(\ell)}$ under ETH.

\begin{figure}[H]
    \centering

\begin{center}

\[
\Qcircuit @C=0.7em @R=1.6em {
\lstick{} & \multigate{2}{\QFTGate} & \ctrl{3} & \qw      & \qw      & \qw    & \multigate{2}{\QFTGate^{-1}} \\
\vdots    & \nghost{\QFTGate}       &          &          &          &        & \nghost{\QFTGate^{-1}}       & & & {  \ket{x+y}}     \\
\lstick{} & \ghost{\QFTGate}        & \qw      & \qw      & \ctrl{3} & \qw    & \ghost{\QFTGate^{-1}}      \inputgroupv{1}{3}{1.2em}{2.6em}{\ket{x}} \gategroup{1}{7}{3}{7}{.8em}{\}} \\
\lstick{} & \multigate{2}{\QFTGate} & \targ    & \qw      & \qw      & \meter & 0  \\
\vdots    & \nghost{\QFTGate}       &          &          &          & \vdots &    \\
\lstick{} & \ghost{\QFTGate}        & \qw      & \qw      & \targ    & \meter & 0      \inputgroupv{4}{6}{1.2em}{2.6em}{\ket{y}} \\
}\]

\end{center}
    \caption{A circuit implementing binary addition using post-selection and three $\QFTGate$ gates.}
    \label{fig:QFT_to_ADD}
\end{figure}
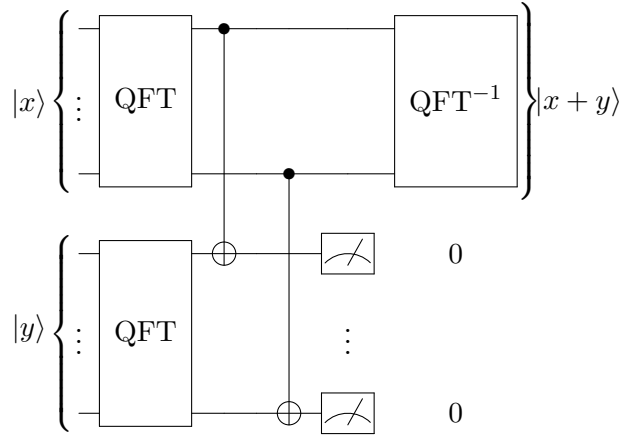

\Paragraph{$\QFTGate_\ell$ gates.}
We now derive a bound on the rank of the $\QFTGate_\ell$ gate, which implements the Quantum Fourier Transform over $\ell$-bit quantum states:
\[
\ket{x}\arrow{\QFTGate_\ell} \sum_{z\in \{0,1\}^{\ell}} e^{2i\pi xz2^{-\ell}}\ket{z}\ .
\]
We implement an $\ADDGate_\ell$ gate via the circuit in \cref{fig:QFT_to_ADD}, which uses three $\QFTGate_\ell$ gates combined with $\ell$ post-selecting measurements.

The circuit performs the following transformation:

\begin{align*}
    \ket{x}\ket{y}  &\arrow{\QFTGate_\ell^{\otimes 2}} \sum_{z\in \{0,1\}^{\ell}} e^{2i\pi xz2^{-\ell}}\ket{z} \sum_{w\in \{0,1\}^{\ell}} e^{2i\pi yw2^{-\ell}}\ket{w} \\
    &\propto \sum_{z,w\in \{0,1\}^{\ell}} e^{2i\pi xz2^{-\ell}} e^{2i\pi yw2^{-\ell}}\ket{z} \ket{w} \\
    &\propto \sum_{z,w\in \{0,1\}^{\ell}} e^{2i\pi (xz+yw)2^{-\ell}}\ket{z} \ket{w} \\
    &\arrow{\CXGate^{\otimes \ell}} \sum_{z,w\in \{0,1\}^{\ell}} e^{2i\pi (xz+y(w\oplus z))2^{-\ell}}\ket{z} \ket{w} \\
    &\arrow{Meas.} \sum_{z\in \{0,1\}^{\ell}} e^{2i\pi (xz+yz)2^{-\ell}} \ket{w} \ket{0^\ell} \\
    &\propto \sum_{z\in \{0,1\}^{\ell}} e^{2i\pi (x+y)z2^{-\ell}} \ket{w} \ket{0^\ell} \\
    &\arrow{\QFTGate^{-1}_\ell} \ket{x+y} \ket{0^\ell}
\end{align*} 
This yields 
$\rank{\ADDGate_\ell}\leq \rank{\QFTGate_\ell}^3$, implying that
$\rank{\QFTGate_\ell}$ is not polynomial under $\PneqNP$, and $\rank{\QFTGate_\ell}=2^{\Omega(\ell)}$ under ETH.

This concludes our derivation of \cref{theorem:hyperpolynomial_rank}. 
\section{Implementation and Experimentation}\label{sec:implementation_and_benchmarking}

We implemented the simulator behind \cref{theorem:main_algorithm} (both the single-threaded and parallelizability results) as a prototype in Python, by extending the phase-sensitive stabilizer simulator of Reardon-Smith~\cite{ReardonSmith2020_PSCS}.
In particular, given a high-level circuit $C$ and a basis state $x$, our simulator performs the following steps.
\begin{compactenum}
\item Gadgetize $C$ to $C'$ using $(\MaxGadgetSize,k)$-decompositions of each non-stabilizer gate, following the process described in \cref{sec:simulation} that re-uses the ancilla qubits that store the individual magic state of each gadget.
\item For each stabilizer state in the stabilizer decomposition, run the phase-sensitive simulator of~\cite{ReardonSmith2020_PSCS} and obtain the coefficient on state $x$.
The individual simulations can run in parallel, for a user-specified number of threads.
\item Maintain a running sum of results that is used to output the final measurement probability.
\end{compactenum}

The rest of this section summarizes an experimental evaluation of this simulation approach of high-level circuits, and compares it against the baseline approach of first compiling to a low-level circuit and then using the standard Qiskit Aer simulator on the low-level circuit.

\Paragraph{Benchmarks.}
We consider benchmarks from four categories: (i)~the state preparation algorithm CVO-QRAM~\cite{de_Veras2022_CVO_QRAM_quantum_algorithm}; (ii)~compositions of minimal oracle functions~\cite{Meuli2018SAT_basedCX_TQuantumCircuitSynthesis}; (iii)~Grover's algorithm~\cite{Grover1996FastQuantumMechanicalAlgorithm}; and (iv)~a string comparison oracle.
All these benchmarks contain high-level gates that operate on a variable number of qubits.

\Paragraph{Low-level circuit simulators.}
For simulating the compiled, low-level circuits, we use the (weak) simulators implemented as part of the Qiskit Aer Python library~\cite{qiskit_aer_2024}. 
These simulators are parameterized by a number of samples to generate from the output distribution.
We set this number to be 1.
Although a single measurement reveals very little information about the measurement probability, we make this choice to ensure that the speedup of our simulator comes from its direct handling of the high-level gates, and not from the fact that it does not have to sample the output distribution.

We use 5 simulation baselines implemented in Aer: matrix product state~\cite{Vidal2003efficient_sim_of_slightly_entangled, Schollwoeck2011matrix_product_state_sim} (MProdS), state vector (SVec), 
density matrix (DMatr), and 
extended stabilizer~\cite{Bravyi2019simulating_q_circuits_by_low_rank_stabilizer_decomp} (ExtStab).
The matrix product state method uses tensors to represent systems of qubits and generally performs well when the amount of entanglement, roughly corresponding to the number of $\CXGate$ gates, is low.
The state vector method represents a pure state of an $n$-qubit system explicitly as a $2^n$-dimensional vector.
This is efficient for systems of few qubits.
The density matrix method stores a mixed state over an $n$-qubit system by representing it as a $2^n \times 2^n$ dimensional matrix.
This is also efficient for systems of few qubits.
Finally, the extended stabilizer method uses magic state injection as our method but only does so for $\TGate$ gates. 
It works well when the number of these non-stabilizer gates is small.

\Paragraph{Experimental setup.}
The experiments are run on an M3 MacBook using 10 GB of RAM and a time-out of 1 hour.
In all our experiments, we compute the probability of measuring $x=0^n$.
Since not all baselines support multithreading, we disable it in our simulator for fair comparison.

\subsection{Experiment: CVO-QRAM}

\begin{figure}[H]
    \centering

\begin{center}

\[
\Qcircuit @C=1em @R=1.6em {
                        &          &          &                  & {\text{Repeat $k$ times}} & & & \\
   \lstick{\ket{1}}     & \ctrl{1} & \ctrl{2} & \ctrl{4}  & \gate{U}  & \ctrl{4} & \ctrl{2} & \ctrl{1} & \qw \\
   \lstick{\ket{0}}     & \targ{}  & \qw      & \qw       & \ctrl{-1} & \qw      & \qw      & \targ{}  & \qw \\
   \lstick{\ket{0}}     & \qw      & \targ{}  & \qw       & \ctrl{-1} & \qw      & \targ{}  & \qw      & \qw \\
                        &          & \vdots   &           &           &          &          &          &     \\
   \lstick{\ket{0}}     & \qw      & \qw      & \targ{}   & \ctrl{-2} & \targ{}  & \qw      & \qw      & \qw \gategroup{2}{2}{6}{8}{1.4em}{--} \\
}\]

\end{center}
    \caption{A circuit describing the structure of the CVO-QRAM experiment.
    In the $j$-th repetition, the repeated section applies a $\CXGate$ gate from the first qubit to any qubit which equals $1$ in $\ket{x_j}$, then applies a $C^n U$ gate, and uncomputes the $\CXGate$ gates.}

    \label{fig:cvo_qram_benchmark}
\end{figure}
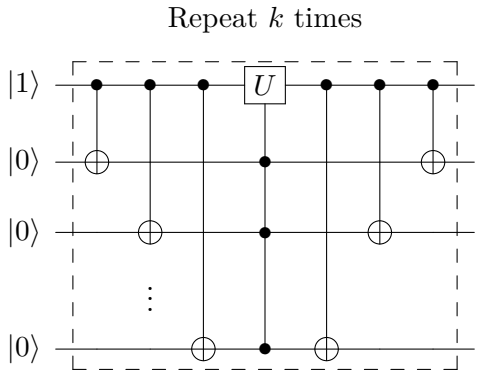

State preparation is an important primitive in quantum software~\cite{Zhang_2022Quantum_State_Preparation}, with active research on designing efficient state preparation circuits~\cite{de_Veras2022_CVO_QRAM_quantum_algorithm,Park_2019Circuit_Based_Quantum_RAM,de_Veras2021_circuit_based_q_RAM}.
As benchmarks, we generate circuits using the CVO-QRAM algorithm~\cite{de_Veras2022_CVO_QRAM_quantum_algorithm}.
This algorithm initializes an $n$-qubit state of the form $\sum_{j=1}^k \psi_j \ket{x_j}$ for basis states $\ket{x_j}$ using $k$ $C^n U$ gates, $O(kn)$ $\CXGate$ gates, and one ancilla qubit.
The structure of the circuits is captured in \cref{fig:cvo_qram_benchmark}.
To obtain the low-level circuits that Aer can handle, we used the Qiskit compiler~\cite{qiskit_aer_2024} to obtain Clifford+$\TGate$ equivalents.
To use our tool, we combined \cref{eq:rank_C_phi_U} of \cref{theorem:gate_rank_from_E_rank} and \cref{eq:rank_str_eq} of \cref{theorem:rank_E_for_comparisons} in order to simulate the $C^n U$ gates.

\begin{table*}[t]
\centering
\begin{tabular}{|rr|l|llll|}
\multicolumn{2}{c}{Benchmark} &  \multicolumn{1}{c}{High-level (direct)} & \multicolumn{4}{c}{Low-level (transpiled)}\\
\hline
    $n$ &   $k$ & Our simulator   & MProdS   & SVec   & DMatr   & ExtStab   \\
\hline
\hline
    5 &      1 & 2ms             & 2ms      & \s{1ms}    & 2ms     & 3.358s     \\
    5 &      2 & 2ms             & \s{1ms}      & 1ms    & 1ms     & 1m15s      \\
    5 &      3 & 379ms           & 6ms      & \s{2ms}    & 5ms     & \OOM       \\
    5 &      4 & 450ms           & \s{6ms}      & 6ms    & 7ms     & \OOM       \\
\hline
   10 &      1 & \s{3ms}         & 7ms      & 4ms        & 4.227s  & \OOM       \\
   10 &      2 & 41ms            & 6ms      & \s{5ms}    & 3.154s  & \OOM       \\
   10 &      3 & 588ms           & 16ms     & \s{13ms}   & 8.534s  & \OOM       \\
   10 &      4 & 7.603s          & 41ms     & \s{32ms}   & 24.16s  & \OOM       \\
\hline
   20 &      1 & \s{6ms}             & 53ms     & 1.778s & \OOM     & \OOM       \\
   20 &      2 & \s{101ms}           & 123ms    & 5.929s & \OOM     & \OOM       \\
   20 &      3 & 1.179s          & \s{150ms}    & 6.138s & \OOM     & \OOM       \\
   20 &      4 & 12.89s          & \s{255ms}    & 13.03s & \OOM     & \OOM       \\
\hline
   50 &      1 & \s{11ms}            & 305ms    & \OOM    & \OOM     & \OOM       \\
   50 &      2 & \s{211ms}           & 811ms    & \OOM    & \OOM     & \OOM       \\
   50 &      3 & 2.689s          & \s{1.265s}   & \OOM    & \OOM     & \OOM       \\
   50 &      4 & 26.88s          & \s{1.602s}   & \OOM    & \OOM     & \OOM       \\
\hline
  500 &      1 & \s{172ms}           & 2m29s    & \OOM    & \OOM     & \OOM       \\
  500 &      2 & \s{2.859s}          & 5m29s    & \OOM    & \OOM     & \OOM       \\
  500 &      3 & \s{37.02s}          & 8m13s    & \OOM    & \OOM     & \OOM       \\
  500 &      4 & \s{6m55s}           & 11m9s    & \OOM    & \OOM     & \OOM       \\
\hline
 1000 &      1 & \s{537ms}           & 24m35s   & \OOM    & \OOM     & \OOM       \\
 1000 &      2 & \s{7.702s}          & 44m42s   & \OOM    & \OOM     & \OOM       \\
 1000 &      3 & \s{1m42s}           & \TO       & \OOM    & \OOM     & \OOM       \\
 1000 &      4 & \s{25m12s}          & \TO       & \OOM    & \OOM     & \OOM       \\
\hline
\end{tabular}

\caption{Experimental results for the CVO-QRAM experiment.
An entry highlighted in green indicates the fastest run-time for that benchmark, while an entry in red indicates either out-of-memory (OOM) or time-out (TO).}

\label{tab:experimental_results_CVO_QRAM}

\end{table*}

\Paragraph{Results.} 
A summary of results is shown in \cref{tab:experimental_results_CVO_QRAM}, for varying numbers of qubits $n$ and $C^n U$ gates $k$.
We observe that our direct simulator is generally faster, and typically by several orders of magnitude.
On the other hand, SVec and MProdS perform efficient simulations on the smaller circuits, but either time out or go out of memory on circuits with many qubits.
DMatr manages to handle circuits of few qubits, but is considerably slower than the above methods, and quickly runs out of memory on larger circuits.
ExtStab quickly runs out of memory, handling only the smallest benchmarks.

\subsection{Experiment: Oracle functions}

In quantum computing, an oracle is a Boolean predicate $\varphi\colon \{0,1\}^k\arrow{} \{0,1\}$, access to which is usually provided via a $C_\varphi \XGate$ gate.
This is a very common construction, used in Grover's algorithm~\cite{Grover1996FastQuantumMechanicalAlgorithm}, the Deutsch-Josza algorithm~\cite{DeutschJozsa1992RapidSolutionProblemsQuantumComputation}, 
the Bernstein-Vazirani algorithm~\cite{BernsteinVazirani1993QuantumComplexityTheory}, and Simon's algorithm~\cite{Simon1997OnThePowerOfQuantumComputation}, among others.

\begin{figure}[H]
    \centering
\begin{center}

\newcommand{\HG}[1]{\gate{\HGate^{\otimes #1}}}
\newcommand{\FG}[1]{\gate{\varphi_{#1}} \qwx[1]}
\newcommand{\MG}[1]{\multigate{1}{\varphi_{#1}}}
\newcommand{\GG}[1]{\ghost{\varphi_{#1}} }

\[
\Qcircuit @C=1em @R=1.6em {
\lstick{\ket{0^5}} & \qw {/} & \HG{5} & \FG{1}  &                 & \\
\lstick{\ket{0}}   & \qw     & \qw    & \targ{} & \MG{2}          & \\
\lstick{\ket{0^4}} & \qw {/} & \HG{4} & \qw     & \GG{2}          &          \\
                   &         &        &         & \vdots          &     \\
\lstick{\ket{0}}   & \qw     & \qw    & \qw     & \targ{}         & \MG{k}             \\
\lstick{\ket{0^4}} & \qw {/} & \HG{4} & \qw     & \qw             & \GG{k} \qwx[1] &            \\
\lstick{\ket{0}}   & \qw     & \qw    & \qw     & \qw             & \targ{}        & \qw  \\
}\]

\end{center}
    \caption{A circuit describing the composed oracle benchmark.}
    \label{fig:oracle_function_benchmark_combined}
\end{figure}
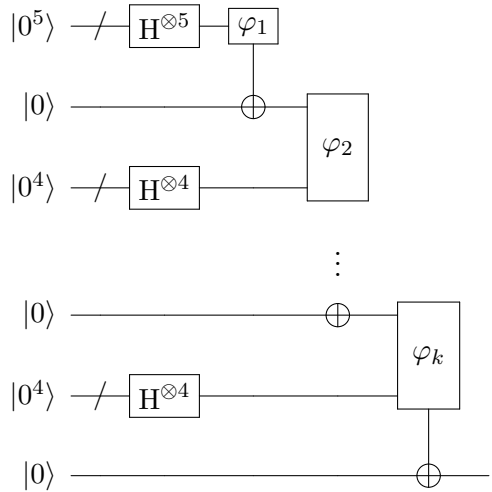

There are 48 spectral equivalent classes over Boolean predicates of the form $\{0,1\}^5\arrow{} \{0,1\}$~\cite{Meuli2018SAT_basedCX_TQuantumCircuitSynthesis}.
These 48 can be composed to give other predicates, and hence their minimal representations as stabilizer+$\TGate$ circuits have been studied~\cite{Meuli2020EnumeratingOptimalQuantumCircuitsUsingSpectralClassification}.
For each of these classes, the optimal stabilizer+$\TGate$ circuit implementing a representative is provided in~\cite{Meuli2020STG_BenchSuite}.
For our experimental setup, we take $k$ representatives at random and chain them together as illustrated in \cref{fig:oracle_function_benchmark_combined}.
To obtain the low-level circuits, we composed the minimal implementations.
To run our simulator, we instead directly implemented the oracle gates using \cref{eq:rank_y_eq_f_of_x_generally} of \cref{theorem:rank_of_effectful_gate} combined with \cref{eq:rank_C_phi_Rz_bound_E,eq:rank_C_phi_Z_eq_rank_C_phi_X} of \cref{theorem:gate_rank_from_E_rank}.

\begin{table*}[t]
\centering

\begin{tabular}{|r|l|llll|}
\multicolumn{1}{c}{Benchmark} &  \multicolumn{1}{c}{High-level (direct)} & \multicolumn{4}{c}{Low-level (transpiled)}\\
\hline
   Oracles $k$ & Direct   & MProdS   & SVec   & DMatr   & ExtStab   \\
\hline
\hline
         1 & 5ms      & \s{1ms}  & \s{1ms}& \s{1ms} & 1.678s    \\
         2 & 16ms     & 6ms      & \s{4ms}& 915ms   & \OOM      \\
         3 & 68ms     & \s{24ms} & 67ms   & \OOM    & \OOM      \\
         4 & \s{340ms}& 4.387s   & 678ms  & \OOM    & \OOM      \\
         5 & \s{3.331s}& 3.406s  & 1m8s   & \OOM    & \OOM      \\
         6 & \s{22.04s}& 1m49s   & \OOM   & \OOM    & \OOM      \\
         7 & \s{54.90s}& \TO     & \OOM   & \OOM    & \OOM      \\
         8 & \s{2m30s} & \TO     & \OOM   & \OOM    & \OOM      \\
         9 & \s{13m31s}& \TO     & \OOM   & \OOM    & \OOM      \\
        10 & \TO       & \TO     & \OOM   & \OOM    & \OOM      \\
\hline
\end{tabular}

\caption{Experimental results for the chained oracle experiment.
An entry highlighted in green indicates the fastest run-time for that benchmark, while an entry in red indicates either out-of-memory (OOM) or time-out (TO).}
\label{tab:experimental_results_oracle_combined}
\end{table*}

\Paragraph{Results.} 
A summary of results is shown in \cref{tab:experimental_results_oracle_combined} for a varying number of chained oracles $k$. 
For non-trivially sized instances, we see that our direct simulator is faster.
MProdS runs out of time on instances of more than 6 oracles.
While SVec and DMatr perform well on small instances, they quickly run out of memory.
ExtStab runs out of memory even on instances of more than one oracle.

\subsection{Experiment: Grover's algorithm}

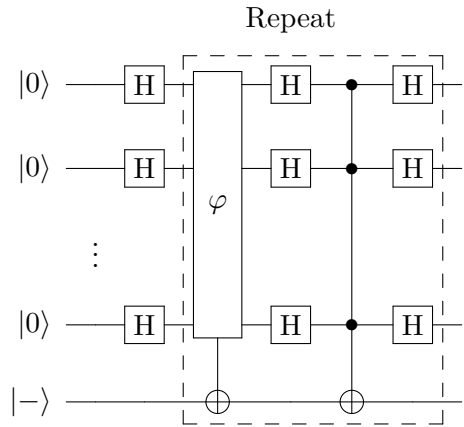
\begin{figure}[H]
    \centering

\begin{center}

\[
\Qcircuit @C=1em @R=1.6em {
                        &        &               &                                  & {\text{Repeat}} & & & \\
   \lstick{\ket{0}}     & \qw    & \gate{\HGate} & \ghost{\varphi}                  & \gate{\HGate} & \ctrl{1} & \gate{\HGate} & \qw\\
   \lstick{\ket{0}}     & \qw    & \gate{\HGate} & \ghost{\varphi}                  & \gate{\HGate} & \ctrl{2}  & \gate{\HGate} & \qw\\
                        & \vdots &               & \nghost{\varphi}                 &               &          &               \\
   \lstick{\ket{0}}     & \qw    & \gate{\HGate} & \multigate{-3}{\varphi} \qwx[1]  & \gate{\HGate} & \ctrl{1} & \gate{\HGate} & \qw \\
   \lstick{\ket{-}}     & \qw    & \qw           & \targ{} & \qw                    & \targ    & \qw & \qw \gategroup{2}{4}{6}{7}{.7em}{--} \\
}\]

\end{center}
    \caption{A circuit describing the structure of the Grover experiment.}
    \label{fig:grover_benchmark}
\end{figure}

Grover's algorithm~\cite{Grover1996FastQuantumMechanicalAlgorithm} is a famous quantum algorithm for finding a satisfying assignment to a propositional formula $\varphi$, offering quadratic speedups over classical approaches.
\cref{fig:grover_benchmark} describes the layout of the circuit implementing the algorithm.

We perform two experiments based on Grover's algorithm, using different choices of $\varphi$.
For the first experiment, we choose $\varphi(x)=\bar{x_1}\wedge \dots \wedge \bar{x_n}$, then generate and run instances for different choices of $n$.
To obtain low-level circuits, we use the standard Qiskit compiler~\cite{qiskit_aer_2024}, while to run ours, we use \cref{eq:rank_C_phi_Z_eq_rank_C_phi_X,eq:C_varphi_Rz_magic_state_decomp} combined with \cref{eq:rank_str_eq}.
For the second experiment, we use $\varphi=\bigwedge_{j=1}^3 \bigvee_{\ell=1}^{n} a_{j,\ell}$, where each $a_{j,\ell}\in \{x_\ell,\bar{x_\ell}\}$ is a literal chosen uniformly at random. 
To obtain low-level circuits, we straightforwardly implement $\varphi$ using three $C^k\XGate$ gates, some $\XGate$ gates, and a $C^3 \XGate$ gate, then compile using the Qiskit compiler~\cite{qiskit_aer_2024}.
To run our simulator, we use \cref{eq:rank_E_conj_bound,eq:rank_E_disj_bound} to implement oracles defined from predicates in conjunctive normal form.

\begin{table*}[t]
\centering

\begin{tabular}{|r|l|llll|}
\multicolumn{1}{c}{Benchmark} &  \multicolumn{1}{c}{High-level (direct)} & \multicolumn{4}{c}{Low-level (transpiled)}\\
\hline
   Size $n$ & Direct       & MProdS    & SVec       & DMatr    & ExtStab    \\
\hline
\hline
      5 & 2.072s       & 17ms      & \s{2ms}    & 7ms      & \OOM       \\
     10 & 3.960s       & 221ms     & \s{2ms}    & 26.68s   & \OOM       \\
     15 & 6.163s       & 1.407s    & \s{7ms}    & \OOM     & \OOM       \\
     20 & 7.979s       & 5.350s    & \s{147ms}  & \OOM     & \OOM       \\
     25 & 9.939s       & 8.545s    & \s{3.586s} & \OOM     & \OOM       \\
     30 & \s{12.01s}   & 14.18s    & \OOM       & \OOM     & \OOM       \\
     40 & \s{15.92s}   & 57.45s    & \OOM       & \OOM     & \OOM       \\
     45 & \s{17.98s}   & 1m40s     & \OOM       & \OOM     & \OOM       \\
     50 & \s{19.92s}   & 2m17s     & \OOM       & \OOM     & \OOM       \\
     75 & \s{30.57s}   & 13m50s    & \OOM       & \OOM     & \OOM       \\
    100 & \s{41.30s}   & \TO       & \OOM       & \OOM     & \OOM       \\
    200 & \s{1m28s}    & \TO       & \OOM       & \OOM     & \OOM       \\
    500 & \s{4m53s}    & \TO       & \OOM       & \OOM     & \OOM       \\
   1000 & \s{13m23s}   & \TO       & \OOM       & \OOM     & \OOM       \\
\hline
\end{tabular}

\caption{Experimental results for simulation of Grover's algorithm with an oracle given a conjunction of negated variables.
An entry highlighted in green indicates the fastest run-time for that benchmark, while an entry in red indicates either out-of-memory (OOM) or time-out (TO).}

\label{tab:experimental_results_grover}

\end{table*}

\Paragraph{Results.}
Our results for the first experiment are captured in \cref{tab:experimental_results_grover}.
For small instances, SVec is faster, but for all instances of 30 qubits or more, our method is more efficient, while from 100 qubits and beyond, our method is the only one that finishes within the bound on memory and time.
Our results for the second experiment are captured in \cref{tab:experimental_results_grover_cnf}.
For small instances, SVec is faster, while from instances of 30 qubits and beyond, our method is the fastest.
From 75 qubits and beyond, our method is the only one that finishes within the bound on memory and time.

\begin{table*}[t]
\centering

\begin{tabular}{|r|l|llll|}
\multicolumn{1}{c}{Benchmark} &  \multicolumn{1}{c}{High-level (direct)} & \multicolumn{4}{c}{Low-level (transpiled)}\\
\hline
   Size $n$ & Direct   & MProdS   & SVec   & DMatr   & ExtStab   \\
\hline
\hline
      5 & 9.222s       & 14ms     & \s{1ms}    & 624ms   & \OOM       \\
     10 & 15.30s       & 184ms    & \s{4ms}    & 39m26s  & \OOM       \\
     15 & 23.29s       & 2.030s   & \s{48ms}   & \OOM     & \OOM       \\
     20 & 30.01s       & 8.158s   & \s{1.271s} & \OOM     & \OOM       \\
     25 & 37.50s       & \s{21.08s}   & 1m19s  & \OOM     & \OOM       \\
     30 & \s{46.02s}   & 52.67s   & \OOM    & \OOM     & \OOM       \\
     40 & \s{58.89s}   & 3m37s    & \OOM    & \OOM     & \OOM       \\
     45 & \s{1m5s}     & 5m44s    & \OOM    & \OOM     & \OOM       \\
     50 & \s{1m11s}    & 12m36s   & \OOM    & \OOM     & \OOM       \\
     75 & \s{1m45s}    & \TO       & \OOM    & \OOM     & \OOM       \\
    100 & \s{2m25s}    & \TO       & \OOM    & \OOM     & \OOM       \\
    200 & \s{5m19s}    & \TO       & \OOM    & \OOM     & \OOM       \\
    500 & \s{16m49s}   & \TO       & \OOM    & \OOM     & \OOM       \\
   1000 & \s{47m25s}   & \TO       & \OOM    & \OOM     & \OOM       \\
\hline
\end{tabular}

\caption{Experimental results for simulation of Grover's algorithm with an oracle given as a function in conjunctive normal form (CNF).
An entry highlighted in green indicates the fastest run-time for that benchmark, while an entry in red indicates either out-of-memory (OOM) or time-out (TO).}

\label{tab:experimental_results_grover_cnf}

\end{table*}

\subsection{Experiment: String Comparison Oracle}

\begin{figure}[H]
    \centering

\begin{center}

\begin{align*}
\Qcircuit @C=1em @R=1.6em {
   \lstick{x\colon \ket{0^k}}     & {/} \qw & \gate{\HGate^{\oplus k}} & \ghost{x>y}                 & \qw \\
   \lstick{y\colon \ket{0^k}}     & {/} \qw & \gate{\HGate^{\oplus k}} & \multigate{-1}{x>y} \qwx[1] & \qw \\
   \lstick{\ket{0}}         &     \qw & \qw                      & \targ                       & \qw \\
}
\end{align*}

\end{center}
    \caption{A circuit describing the structure of the $C_{x>y}\XGate$ benchmarks.}
    \label{fig:benchmark_x_gt_y}
\end{figure}
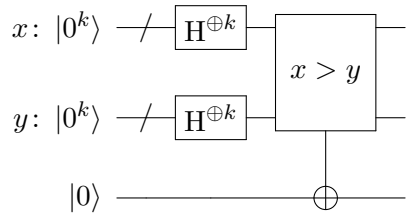

Our final experiment involves simulating an oracle gate for the Boolean predicate $x>y$, defined for $k$-bit integers $x$ and $y$.
The oracle is provided as a $C_{x>y}\XGate$ gate.
For different choices of $k$, we simulate comparing a uniform superposition over all choices of $x$ and $y$, as described in \cref{fig:benchmark_x_gt_y}.

To obtain low-level circuits, we use the implementation provided in~\cite{Li2012Design_quantum_comparator} to compile to Clifford+$C^k \XGate$ circuits, then compile to Clifford+$\TGate$ using the Qiskit compiler~\cite{qiskit_aer_2024}.
Their implementation uses $k$ ancillary bits to store intermediate computations.
Hence in \cref{tab:experiment_GT_gate}, we have $n=3k+1$, as we need $2k+1$ qubits to store $x$, $y$, and the target qubit.
To run our simulator, we use \cref{eq:rank_C_phi_Z_eq_rank_C_phi_X,eq:C_varphi_Rz_magic_state_decomp} combined with \cref{eq:rank_str_gt} to simulate directly, while the results in the ``As $C^k \XGate$'' column are produced by instead running the Clifford+$C^k \XGate$ circuits.

\begin{table*}[t]
\centering
\begin{tabular}{|rr|ll|llll|}
\multicolumn{2}{c}{Benchmark} &  \multicolumn{2}{c}{High-level (direct)} & \multicolumn{4}{c}{Low-level (transpiled)}\\
\hline
$k$   & $n$ & Direct   & As $C^k \XGate$ & MProdS   & SVec      & DMatr    & ExtStab  \\
\hline
\hline
1     & 4   & 0ms      & 0ms            & 1ms      & \s{0ms}   & 1ms      & 107ms    \\
2     & 7   & 0ms      & 9ms            & 1ms      & \s{0ms}   & 1ms      & 314ms    \\
3     & 10  & 1ms      & 117ms          & 3ms      & \s{0ms}   & 238ms    & 4m47s    \\
4     & 13  & 2ms      & 1.313s         & 8ms      & \s{1ms}   & 32.13s   & \OOM     \\
5     & 16  & \s{2ms}  & 13.78s         & 22ms     & 3ms       & \OOM     & \OOM     \\
6     & 19  & \s{3ms}  & 2m22s          & 81ms     & 17ms      & \OOM     & \OOM     \\
7     & 22  & \s{4ms}  & 22m56s         & 646ms    & 116ms     & \OOM     & \OOM     \\
8     & 25  & \s{5ms}  & \TO            & 3m47s    & 985ms     & \OOM     & \OOM     \\
9     & 28  & \s{6ms}  & \TO            & 2m55s    & 8.514s    & \OOM     & \OOM     \\
10    & 31  & \s{9ms}  & \TO            & 37m52s   & \OOM      & \OOM     & \OOM     \\
11    & 34  & \s{9ms}  & \TO            & \TO      & \OOM      & \OOM     & \OOM     \\
12    & 37  & \s{19ms} & \TO            & \TO      & \OOM      & \OOM     & \OOM     \\
13    & 40  & \s{17ms} & \TO            & \TO      & \OOM      & \OOM     & \OOM     \\
14    & 43  & \s{17ms} & \TO            & \TO      & \OOM      & \OOM     & \OOM     \\
15    & 46  & \s{18ms} & \TO            & \TO      & \OOM      & \OOM     & \OOM     \\
\hline
\end{tabular}

\caption{Experimental results for simulating a $C_{x>y} \XGate$ gate.
An entry highlighted in green indicates the fastest run-time for that experiment, while an entry in red indicates either out-of-memory (OOM) or time-out (TO).}

\label{tab:experiment_GT_gate}

\end{table*}

\Paragraph{Results.} 
See \cref{tab:experiment_GT_gate} for results.
Between our results, we see that the direct simulation outperforms simulating Clifford+$C^k \XGate$ circuits on all benchmark sizes.
Comparing against Qiskit Aer, SVec is generally faster for small instances, while on instances of 16 qubits and beyond, our direct simulation results are faster. 
Simulating the Clifford+$C^k \XGate$ circuits, while consistently outperforming DMatr and ExtStab, is itself consistently outperformed by MProdS and SVec, suggesting increased benefits when working with higher gate-levels.
On instances of 34 qubits or more, our direct method is the only one which does not go out of time or memory.

\section{Related Work}

The ZX calculus can be applied to perform strong simulation of universal quantum circuits.
The work of~\cite{Kissinger2022classical_simulation_partial_and_graphical_stabilizer_decomposition} converts Clifford+$\TGate$ and Clifford+$\CCZGate$ circuits to ZX diagram representations, then recursively apply diagram simplifications and magic state decompositions to simulate these efficiently. 
These decompositions depend on the rank of the magic state $\ket{\TGate}^{\otimes t}$, which is shown to be $\rank{\ket{\TGate}^{\otimes t}} \leq 2^{0.396 t}$, improving the earlier result of~\cite{Gosset2021improved_upper_bounds_stabilizer_rank} which showed that this upper bound holds when $t$ is sufficiently large.
However, dealing with arbitrary high-level gates, this technique still requires a compilation step, typically introducing extra non-Clifford gates.
In contrast, our technique sidesteps this overhead by providing magic state decompositions of high-level gates directly.

The work of~\cite{Koch_2024} expands on the technique of~\cite{Kissinger2022classical_simulation_partial_and_graphical_stabilizer_decomposition} by using so-called ``triangle nodes'' in the ZX diagrams. 
This permits an efficient decomposition of $C^k \XGate$ gates, resulting in a $2^t$ exponential factor for $t$ $C^k X$ gates.
While our work also achieves a $2^t$ exponential factor for $t$ $C^k \XGate$ gates, we generalize to arbitrary oracle gates beyond a single conjunction.

The work of~\cite{ahmad2024dynamictdecompositionclassicalsimulation} further extends the idea by considering a number of decompositions which apply to substructures of the ZX diagram and permit efficient decomposition when they are present.
Heuristics determine which decomposition to apply, for which experiments seem to suggest practical improvements over a greedy strategy.
The work of~\cite{sutcliffe2024smarterkpartitioningzxdiagramsimproved} studies partitioning the ZX diagram into smaller diagrams which can be individually reduced, and from which the simulation result can be derived. 
The efficacy each technique depends on the structure of the underlying graph, for which our technique is oblivious.
Future work could explore a heuristic interleaving of partitions and decompositions to improve efficiency.

The recent work of~\cite{Kuyanov2026efficientclassicalsimulationlowrankwidth} gives formal bounds on the run-time of simulation based on the ``rank-width'' $R$ of the ZX diagram, giving a run-time of $\Tilde{O}(4^R)$, where $\Tilde{O}$ hides sub-exponential factors.
The recent work of~\cite{Codsi2026unifying} unifies approaches from stabilizer decompositions and tensor network contraction to give run-times based on the structure of the underlying graph.
For a stabilizer+$\TGate$ circuit $C$ with rank-width $\texttt{rw}(C)$ and tree-width $\texttt{tw}(C)$, they give algorithms with respective run-times of $\Tilde{O}(t^{\gamma \cdot \texttt{rw}(C)})$ and $\Tilde{O}(t^{\texttt{tw}(C)})$, where $t$ is the number of $\TGate$ gates of $C$, and $\gamma \approx 3.42$.
Future work could generalize their work to our approach for high-level gates, by exploring the graph structure of $C$ to improve efficiency.

Finally, the work of~\cite{qassim2019clifford} proposes a technique for improving the simulation of the Clifford fragments of a general circuit.
In high level, this approach recompiles the input circuits to permit efficient simulation via a sum-over-Cliffords technique.
It is possible to apply a similar recompilation strategy to our work: Given a circuit $C$, write it as $C_t \GATE_t C_{t-1} \GATE_{t-1} \dots C_1 \GATE_1 C_0$ for Clifford circuits $C_j$ and non-Clifford gates $\GATE_j$, then apply standard techniques to canonicalize the Clifford circuits to $O(n^2)$ gates in $O(nm + n^3)$ time.
This brings the circuit size to $O(n^2 t)$, resulting in a run-time of $O(\chi n^2 (n^2 t + t\mu) + nm + n^3) = O(\chi n^4 t+nm)$, where we use the fact that $\mu$ is the size of a Clifford circuit and hence bounded as $O(n^2)$.
This strategy outperforms the baseline algorithm of~\cref{theorem:main_algorithm} when $m=\omega(n^2 t)$.
\section{Conclusion}\label{sec:conclusion}

We have presented a gadget-based simulator which gadgetizes, then simulates high-level gates directly using magic state injection, thereby avoiding the blowup of compilation.
We have also shown that a number of high-level gates that are often used in quantum computing enjoy a small stabilizer rank (in particular, independent of the number of qubits they act on), and can therefore be simulated more efficiently using our gadget-based simulator.
Our experiments with various high-level gates indeed show reduced simulation time for several types of high-level circuits compared to Qiskit Aer that simulates via compilation.
Exploring the limits of our technique, we have shown that the stabilizer rank of certain high-level gates is hyper-polynomial under $\PneqNP$, and exponential under ETH.

For each of the high-level gates which we have considered here, reducing the upper bounds on the rank of the corresponding magic states reduces the overall simulation complexity for circuits containing that gate.
To complement such results, corresponding lower-bounds on the magic state rank help guide the search for efficient simulation, by avoiding searching for improvements where they are unlikely to be found.
Furthermore, extending the high-level gate simulation theory established here to other gates, more circuits might be simulated efficiently.
With a tool for direct strong simulation of high-level gates, circuit equivalence can be decided without suffering the blow-up of compilation.
If applied efficiently, this could extend the reach of formal quantum circuit verification.

\section{Acknowledgments}

This work was partially supported by a research grant (VIL42117) from VILLUM FONDEN.

\bibliographystyle{quantum} 
\bibliography{references}

\end{document}